\documentclass[pre,twocolumn,showpacs,showkeys,preprintnumbers,amsmath,amssymb]{revtex4}

\usepackage{graphicx}
\usepackage{dcolumn}
\usepackage{bm}
\usepackage{subfigure}
\usepackage{amssymb}
 \usepackage{graphics}
  \usepackage{epsfig}

\begin{document}


\title{An introduced effective-field theory study of spin-1 transverse Ising model with crystal field anisotropy in a longitudinal magnetic field }

\author{Yusuf Y\"{u}ksel}
\author{Hamza Polat}\email{hamza.polat@deu.edu.tr}

\affiliation{ Department of Physics, Dokuz Eyl\"{u}l University,
TR-35160 Izmir, Turkey} \affiliation{ Dokuz Eyl\"{u}l University,
Graduate School of Natural and Applied Sciences}
\date{\today}

\begin{abstract}
A spin-1 transverse Ising model with longitudinal crystal field in a
longitudinal magnetic field is examined by introducing an effective
field approximation (IEFT) which includes the correlations between
different spins that emerge when expanding the identities. The
effects of the crystal field as well as the transverse and
longitudinal magnetic fields on the thermal and magnetic properties
of the spin system are discussed in detail. The order parameters,
Helmholtz free energy and entropy curves are calculated numerically
as functions of the temperature and Hamiltonian parameters. A number
of interesting phenomena such as reentrant phenomena originating
from the temperature, crystal field, transverse and longitudinal
magnetic fields have been found.
\end{abstract}



\maketitle

\section{Introduction} \label{introduction}
Ising model in a transverse field has been widely examined in
statistical mechanics and condensed matter physics since the
pioneering work of de Gennes \cite{deGennes} who introduced it as a
pseudo spin model for hydrogen-bonded ferroelectrics such as the
$KH_{2}PO_{4}$ type. Following studies has been predicated that this
semi-quantum mechanical model can be successfully applied to a
variety of physical systems such as $DyVO_{4}$, $TbVO_{4}$
\cite{elliot} and some real magnetic materials \cite{wong}. From the
theoretical point of view, transverse Ising model (TIM) has been
investigated by a variety of techniques such as renormalization
group method (RG) \cite{fisher_rg}, effective field theory (EFT)
\cite{saber_eft,sarmento_eft,Saber}, cluster variation method (CVM)
\cite{saxena_cvm}, mean field theory (MFT) \cite{saxena_mf}, pair
approximation (PA) \cite{canko_pa} and Monte Carlo simulations (MC)
\cite{creswick_mc}. In the previous works mentioned above, the
authors focused their attention on the behavior of tricritical
points but have not considered the effect of the crystal field (i.e.
single ion anisotropy) in Hamiltonian describing the system.

However, there are a few studies in the literature that include the
crystal field as well as the transverse field interactions.
Recently, the effect of both the transverse field and the crystal
field on the spin-$S$ Ising model with spins of magnitude $S$=1 have
been studied and it is shown that TIM model presents a rich variety
of critical phenomena. For example, Jiang et al. \cite{Jiang1} has
studied the spin-1 TIM on a honeycomb lattice with a longitudinal
crystal field and discussed the existence of a tricritical point at
which the phase transition change from second-order to first order.
By using the EFT with a probability distribution technique, Htoutou
et al. \cite{Htoutou2} have investigated the influence of the
crystal field on the phase diagrams of a site diluted spin-1 TIM on
a square lattice. Similarly, Jiang have studied a bond diluted
spin-1 TIM with crystal field interaction for a honeycomb lattice
within the framework of the EFT with correlations \cite{Jiang2}. In
these studies, the authors have reported the observation of a
reentrant behavior on the system. Furthermore, in a series of papers
Htoutou et al. \cite{Htoutou1,Htoutou3} have discussed the
dependence of the behavior of the order parameters on the transverse
and crystal fields, but they have restricted themselves on the
second order transition properties. The effect of a longitudinal
crystal field on the phase transitions in spin-$3/2$ and spin-$2$
transverse Ising model has been also examined for both honeycomb and
square lattices by using the EFT with correlations
\cite{wei_jiang1,wei_jiang2}. More recently, within the basis of EFT
and MFT, Miao et al. \cite{Miao} have studied the phase diagrams of
a spin-1 transverse Ising model for a honeycomb lattice. They have
obtained the first-order transition lines by comparing the Gibbs
free energy.

An ordinary EFT approximation includes spin-spin correlations
resulting from the usage of the Van der Waerden identities and
provides results that are much superior to those obtained within the
traditional MFT. However, the EFT approximations mentioned above are
not sufficient enough to improve the results much. The reason may be
due to the usage of a decoupling approximation that neglects the
correlations between different spins that emerge when expanding the
identities. According to us, the first-order transition lines
obtained in Refs. \cite{Jiang1,Htoutou2,Jiang2,Htoutou1, Htoutou3}
and \cite{Miao} are incomplete, because the spin correlation
functions such as $\langle S_{0}S_{1}\rangle$, $\langle
S_{0}S_{1}S_{2}^{2}\rangle$ etc. can not be determined by using any
decoupling approximation or in other words, by neglecting the
correlations between different spins. In order to overcome this
point, we proposed the IEFT approximation that takes into account
the correlations between different spins in the cluster of
considered lattice $(q=3)$ \cite{polat,canpolat,yuksel}. Namely, the
hallmark of the IEFT is to consider the correlations between
different spins that emerge when expanding the identities and this
method is superior to conventional mean field theory and the other
EFT approximations in the literature. Therefore, it is expected that
the calculation results will be more accurate.

As far as we know, there is not such a study which includes the
longitudinal component of the magnetic field in addition to the
crystal field and transverse magnetic field on the Hamiltonian.
Thus, in the present work, we intended to investigate the thermal
and magnetic properties of spin-1 TIM with crystal field under a
longitudinal magnetic field on a honeycomb lattice within the
framework of the IEFT. For this purpose, we investigated the proper
phase diagrams, especially the first-order transition lines that
include reentrant phase transition regions and we improved the
results in Refs.\cite{Jiang1,Htoutou1}. We gave the numerical
results for the behavior of the order parameters when the system
undergoes a first or second order transition at a finite
temperature. In addition, it would be interesting to see how the
thermodynamic properties like entropy $S$ which has not been
calculated before and Helmholtz free energy $F$ are effected by the
crystal field as well as transverse and longitudinal magnetic
fields. Hence, the numerical results are presented and compared with
the literature.

The layout of this paper is as follows. In section
\ref{formulation}, we briefly present the formulations of the IEFT.
The results and discussions are presented in section \ref{results}.
Finally, section \ref{conclusion} contains our conclusions.

\section{Formulation} \label{formulation}
As our model we consider a two dimensional lattice which has $N$
identical spins arranged. We define a cluster on the lattice which
consists a central spin labeled $S_{0}$, and $q$ perimeter spins
being the nearest-neighbors of the central spin. The cluster
consists of $(q+1)$ spins being independent from the value of $S$.
The nearest-neighbor spins are in an effective field produced by the
outer spins, which can be determined by the condition that the
thermal average of the central spin is equal to that of its
nearest-neighbor spins. The Hamiltonian of the spin-1 transverse
model with crystal field in a longitudinal magnetic field is given
by
\begin{equation}\label{eq1}
H=-J\sum_{<i,j>}S_{i}^{z}S_{j}^{z}-D\sum_{i}(S_{i}^{z})^{2}-\Omega\sum_{i}S_{i}^{x}
-h\sum_{i}S_{i}^{z},
\end{equation}
where $S_{i}^z$ and $S_{i}^x$ denote the $z$ and $x$ components of
the spin operator, respectively. The first summation in equation
(\ref{eq1}) is over the nearest-neighbor pairs of spins and the
operator $S_{i}^z$ takes the values $S_{i}^z=0,\pm1$. $J$, $D$,
$\Omega$ and $h$ terms stand for the exchange interaction,
single-ion anisotropy (i.e. crystal field) and transverse and
longitudinal magnetic fields, respectively.

At first, we start constructing the mathematical background of our
model by using the approximated spin correlation identities
introduced by S\'{a}Barreto, Fittipaldi and Zeks \cite{Barreto}
\begin{equation}\label{eq2}
\langle \{f_{i}\}S_{i}^\alpha\rangle=\left\langle \{f_{i}\}
\frac{Tr_{i}S_{i}^\alpha\exp{(-\beta H_{i})}}{Tr_{i}\exp{(-\beta
H_{i})}}\right\rangle,
\end{equation}
\begin{equation}\label{eq3}
\langle \{f_{i}\}(S_{i}^\alpha)^2\rangle=\left\langle \{f_{i}\}
\frac{Tr_{i}(S_{i}^\alpha)^2\exp{(-\beta H_{i})}}{Tr_{i}\exp{(-\beta
H_{i})}}\right\rangle,
\end{equation}
where $\beta=1/k_{B}T$ and  $\alpha=z$ or $x$.

In order to apply the differential operator technique, we should
separate the Hamiltonian (\ref{eq1}) into two parts as
$H=H_{i}+H^{'}$. Here, one part denoted by $H_{i}$ includes all
contributions associated with the site $i$, and the other part
$H^{'}$ does not depend on the site $i$. At this point, one should
notice that $H_{i}$ and $H^{'}$ do not commute with each other. We
can write $-H_{i}$ as
\begin{equation}\label{eq4}
-H_{i}=E_{i}S_{i}^{z}+D\left(S_{i}^{z}\right)^{2}+\Omega
S_{i}^{x}+hS_{i}^{z},
\end{equation}
where $E_{i}=J\sum_{j}S_{j}^z$ is the local field on the site $i$.
If we use the matrix representations of the operators $S_{i}^{z}$
and $S_{i}^{x}$ for the spin-1 system then we can obtain the matrix
form of equation (\ref{eq4})
\begin{equation}\label{eq5}
-H_{i}=\left(
  \begin{array}{ccc}
    E_{i}+D+h & \Omega/\sqrt{2} & 0 \\
    \Omega/\sqrt{2} & 0 & \Omega/\sqrt{2} \\
    0 & \Omega/\sqrt{2} & -E_{i}+D-h \\
  \end{array}
\right).
\end{equation}
In order to proceed further, we have to diagonalize $-H_{i}$ matrix
in equation (\ref{eq5}). The three eigenvalues are
\begin{eqnarray}\label{eq6}
 \nonumber
  \lambda_{1} &=& \frac{2c}{3}+\frac{2p}{3}\cos\left(\frac{\theta}{3}\right), \\
   \lambda_{2}&=&\frac{2c}{3}-\frac{2p}{3}\cos\left(\frac{\pi-\theta}{3}\right),  \\
\nonumber
   \lambda_{3}&=&\frac{2c}{3}-\frac{2p}{3}\cos\left(\frac{\pi+\theta}{3}\right),
\end{eqnarray}
where
\begin{eqnarray*}
   \theta &=& \arccos\left(\frac{\zeta}{p^{3}}\right), \\
   \zeta&=&D\left(9E_{i}^{2}-\frac{9}{2}\Omega^{2}-D^{2}+18E_{i}h+9h^{2}\right), \\
   p^{2}&=&3E_{i}^{2}+3\Omega^{2}+D^{2}+6E_{i}h+3h^{2},
\end{eqnarray*}
and the eigenvectors $\varphi_{k}$ of $-H_{i}$ corresponding to the
eigenvalues in equation (\ref{eq6}) are calculated as follows
\begin{eqnarray}
  \nonumber\alpha_{k}&=&\pm\sqrt{1-\beta_{k}^{2}-\gamma_{k}^{2}},\\
  \nonumber\beta_{k} &=&-\frac{\left[E_{i}+(D+h-\lambda_{k})\right]}{\Omega/\sqrt{2}}\alpha_{k},  \\
  \nonumber\gamma_{k}
  &=&-\frac{\left[E_{i}+(D+h-\lambda_{k})\right]}{\left[E_{i}-(D-h-\lambda_{k})\right]}\alpha_{k},
\end{eqnarray}
\begin{equation}\label{eq7}
\varphi_{k}=\left(
              \begin{array}{c}
                \alpha_{k} \\
                \beta_{k} \\
                \gamma_{k} \\
              \end{array}
            \right),
            \qquad k=1,2,3.
\end{equation}
Hereafter, we apply the differential operator technique in equations
(\ref{eq2}) and (\ref{eq3}) with $\{f_{i}\}=1$. From equation
(\ref{eq2}) we obtain the following spin correlations for the
thermal average of a central spin for honeycomb lattice $(q=3)$ as
\begin{eqnarray}\label{eq8} \langle
\nonumber S_{0}^{z}\rangle&=&\left\langle\prod_{j=1}^{q=3}\left[1+S_{j}^{z}\mathrm{sinh}(J\nabla)+(S_{j}^{z})^{2}\{\mathrm{cosh}(J\nabla)-1\}\right]\right\rangle\\
&&\times F(x)|_{x=0},
\end{eqnarray}
\begin{eqnarray}\label{eq9}
\langle \nonumber
S_{0}^{x}\rangle&=&\left\langle\prod_{j=1}^{q=3}\left[1+S_{j}^{z}\mathrm{sinh}(J\nabla)+(S_{j}^{z})^{2}\{\mathrm{cosh}(J\nabla)-1\}\right]\right\rangle\\
&&\times H(x)|_{x=0}.
\end{eqnarray}
By expanding the right-hand sides of equations (\ref{eq8}) and
(\ref{eq9}) we get the longitudinal and transverse spin correlations
as
\begin{eqnarray}\label{eq10}
\nonumber m_{z}=\langle S_{0}^{z}\rangle&=&l_{0}+3k_{1}\langle S_{1}
\rangle+3(l_{1}-l_{0})\langle S_{1}^{2} \rangle+3l_{2}\langle
S_{1}S_{2}
\rangle \\
\nonumber& & +6(k_{2}-k_{1})\langle S_{1}S_{2}^{2}
\rangle+3(l_{0}-2l_{1}+l_{3})\langle S_{1}^{2}S_{2}^{2}
\rangle \\
\nonumber& &+k_{3}\langle S_{1}S_{2}S_{3}
\rangle+3(l_{4}-l_{2})\langle S_{1}S_{2}S_{3}^{2}
\rangle\\
\nonumber& &+3(k_{1}-2k_{2}+k_{4})\langle S_{1}S_{2}^{2}S_{3}^{2}
\rangle\\
& &+(-l_{0}+3l_{1}-3l_{3}+l_{5})\langle S_{1}^{2}S_{2}^{2}S_{3}^{2}
\rangle,
\end{eqnarray}
\begin{eqnarray}\label{eq11}
\nonumber m_{x}=\langle S_{0}^{x}\rangle&=&p_{0}+3c_{1}\langle S_{1}
\rangle+3(p_{1}-p_{0})\langle S_{1}^{2} \rangle+3p_{2}\langle
S_{1}S_{2}
\rangle \\
\nonumber& & +6(c_{2}-c_{1})\langle S_{1}S_{2}^{2}
\rangle+3(p_{0}-2p_{1}+p_{3})\langle S_{1}^{2}S_{2}^{2}
\rangle \\
\nonumber& &+c_{3}\langle S_{1}S_{2}S_{3}
\rangle+3(p_{4}-p_{2})\langle S_{1}S_{2}S_{3}^{2}
\rangle\\
\nonumber& &+3(c_{1}-2c_{2}+c_{4})\langle S_{1}S_{2}^{2}S_{3}^{2}
\rangle\\
& &+(-p_{0}+3p_{1}-3p_{3}+p_{5})\langle S_{1}^{2}S_{2}^{2}S_{3}^{2}
\rangle.
\end{eqnarray}
Next, the average value of a perimeter spin in the system can be
written as follows and it is found as
\begin{eqnarray}\label{eq12}
\nonumber m_1&=&\langle S_{\delta}^{z}\rangle=\langle
1+S_{0}^{z}\mathrm{sinh}(J\nabla)+(S_{0}^{z})^{2}\{\mathrm{cosh}(J\nabla)-1\}\rangle\\
&&\times F(x+\gamma)|_{x=0},
\end{eqnarray}
\begin{equation}\label{eq13}
\langle S_{1}\rangle=a_{1}\left(1-\langle
(S_{0}^{z})^{2}\rangle\right)+a_{2}\langle S_{0}^{z}
\rangle+a_{3}\langle (S_{0}^{z})^{2}\rangle,
\end{equation}
where $\gamma=(q-1)A$ is the effective field produced by the $(q-1)$
spins outside the system and $A$ is an unknown parameter to be
determined self-consistently. In the effective-field approximation,
the number of independent spin variables describes the considered
system. This number is given by the relation $\nu=\langle
(S_{i}^{z})^{2S}\rangle$. As an example for the spin-1 system,
$2S=2$ which means that we have to introduce the additional
parameters $\langle (S_{0}^{z})^{2}\rangle$, $\langle
(S_{0}^{x})^{2}\rangle$ and $\langle (S_{\delta}^{z})^{2}\rangle$
resulting from the usage of the Van der Waerden identity for the
spin-1 Ising system. With the help of equation (\ref{eq3})
\begin{eqnarray}\label{eq14}
\nonumber q_{z}&=&\langle(S_{0}^{z})^{2}\rangle\\
\nonumber
&=&\left\langle\prod_{j=1}^{q}\left[1+S_{j}^{z}\mathrm{sinh}(J\nabla)
+(S_{j}^{z})^{2}\{\mathrm{cosh}(J\nabla)-1\}\right]\right\rangle\\
&&\times G(x)|_{x=0},
\end{eqnarray}
\begin{eqnarray}\label{eq15}
\nonumber q_{x}&=&\langle(S_{0}^{x})^2\rangle\\
\nonumber&=&\left\langle\prod_{j=1}^{q}\left[1+S_{j}^{z}\mathrm{sinh}(J\nabla)+(S_{j}^{z})^{2}\{\mathrm{cosh}(J\nabla)-1\}\right]\right\rangle\\
&&\times K(x)|_{x=0}.
\end{eqnarray}
Hence, we get the quadrupolar moments by expanding the right-hand
sides of equations (\ref{eq14}) and (\ref{eq15})
\begin{eqnarray}\label{eq16}
\nonumber  \langle (S_{0}^{z})^{2}\rangle &=& r_{0}+3n_{1}\langle S_{1}\rangle+3(r_{1}-r_{0})\langle S_{1}^{2}\rangle+3r_{2}\langle S_{1}S_{2}\rangle \\
\nonumber   &&+6(n_{2}-n_{1})\langle S_{1}S_{2}^{2}\rangle+3(r_{0}-2r_{1}+r_{3})\langle S_{1}^{2}S_{2}^{2}\rangle\\
\nonumber   &&+n_{3}\langle S_{1}S_{2}S_{3}\rangle+ 3(r_{4}-r_{2})\langle S_{1}S_{2}S_{3}^{2}\rangle \\
\nonumber   &&+3(n_{1}-2n_{2}+n_{4})\langle
S_{1}S_{2}^{2}S_{3}^{2}\rangle\\
   &&+(-r_{0}+3r_{1}-3r_{3}+r_{5})\langle
   S_{1}^{2}S_{2}^{2}S_{3}^{2}\rangle,
\end{eqnarray}

\begin{eqnarray}\label{eq17}
\nonumber \langle (S_{0}^{x})^2\rangle&=&v_{0}+3\mu_{1}\langle S_{1}
\rangle+3(v_{1}-v_{0})\langle S_{1}^{2} \rangle+3v_{2}\langle
S_{1}S_{2}
\rangle \\
\nonumber& & +6(\mu_{2}-\mu_{1})\langle S_{1}S_{2}^{2}
\rangle+3(v_{0}-2v_{1}+v_{3})\langle S_{1}^{2}S_{2}^{2}
\rangle \\
\nonumber& &+\mu_{3}\langle S_{1}S_{2}S_{3}
\rangle+3(v_{4}-v_{2})\langle S_{1}S_{2}S_{3}^{2}
\rangle\\
&&\nonumber+3(\mu_{1}-2\mu_{2}+\mu_{4})\langle
S_{1}S_{2}^{2}S_{3}^{2} \rangle\\ &
&+(-v_{0}+3v_{1}-3v_{3}+v_{5})\langle S_{1}^{2}S_{2}^{2}S_{3}^{2}
\rangle.
\end{eqnarray}
Corresponding to equation (\ref{eq12})
\begin{equation}\label{eq18} \langle
(S_{\delta}^{z})^{2}\rangle=\langle
1+S_{0}^{z}\mathrm{sinh}(J\nabla)+(S_{0}^{z})^{2}\{\mathrm{cosh}(J\nabla)-1\}\rangle
G(x+\gamma),
\end{equation}
\begin{equation}\label{eq19}
\langle S_{1}^{2}\rangle=b_{1}\left(1-\langle
(S_{0}^{z})^{2}\rangle\right)+b_{2}\langle S_{0}^{z}
\rangle+b_{3}\langle (S_{0}^{z})^{2}\rangle.
\end{equation}
Details of calculation through (\ref{eq8}-\ref{eq19}) can be found
in Appendix section. With the help of equations (\ref{eq6}) and
(\ref{eq7}) the functions $F(x)$, $G(x)$, $H(x)$ and $K(x)$ in
equations (\ref{eq8}), (\ref{eq9}), (\ref{eq14}) and (\ref{eq15})
can be calculated numerically from the relations
\begin{equation}\label{eq20}
F(x)=\frac{1}{\sum_{n=1}^{s=3}\exp{(\beta\lambda_{n})}}\sum_{n=1}^{s=3}\langle
\varphi_{n}|S_{i}^{z}|\varphi_{n}\rangle\exp{(\beta\lambda_{n})},
\end{equation}
\begin{equation}\label{eq21}
H(x)=\frac{1}{\sum_{n=1}^{s=3}\exp{(\beta\lambda_{n})}}\sum_{n=1}^{s=3}\langle
\varphi_{n}|S_{i}^{x}|\varphi_{n}\rangle\exp{(\beta\lambda_{n})},
\end{equation}
\begin{equation}\label{eq22}
G(x)=\frac{1}{\sum_{n=1}^{s=3}\exp{(\beta\lambda_{n})}}\sum_{n=1}^{s=3}\langle
\varphi_{n}|(S_{i}^{z})^{2}|\varphi_{n}\rangle\exp{(\beta\lambda_{n})},
\end{equation}
\begin{equation}\label{eq23}
K(x)=\frac{1}{\sum_{n=1}^{s=3}\exp{(\beta\lambda_{n})}}\sum_{n=1}^{s=3}\langle
\varphi_{n}|(S_{i}^{x})^2|\varphi_{n}\rangle\exp{(\beta\lambda_{n})}.
\end{equation}
The internal energy $U$ per site of the system can be obtained
easily from the thermal average of the Hamiltonian in equation
(\ref{eq1}). Thus, the internal energy is given by
\begin{equation}\label{eq26}
-\frac{U}{NJ}=\frac{q}{2}\langle S_{0}S_{1}\rangle+D\langle
(S_{0}^z)^{2}\rangle+\Omega\langle S_{0}^{x}\rangle+h\langle
S_{0}^z\rangle,
\end{equation}
where the correlation functions $\langle S_{0}S_{1}\rangle$,
$\langle (S_{0}^z)^{2}\rangle$, $\langle S_{0}^{x}\rangle$ and
$\langle S_{0}^z\rangle$ are obtained from equation (\ref{eqA7}).
With the use of equation (\ref{eq26}), the specific heat of the
system can be numerically determined from the relation
\begin{equation}\label{eq27}
C_{h}=\left(\frac{\partial U}{\partial T}\right)_{h}.
\end{equation}
The Helmholtz free energy of a system is defined as
\begin{equation}\label{eq28}
F=U-TS,
\end{equation}
which, according to the third law, can be written in the form
\cite{Huang}
\begin{equation}\label{eq29}
F=U-T\int_{0}^{T}\frac{C}{T^{'}}dT^{'},
\end{equation}
where the integral in the second term is entropy of the system
according to the second law. First, we calculate the internal energy
per site from equation (\ref{eq26}). Then with the help of equation
(\ref{eq27}) we can carry out numerical integration and calculate
the entropy and free energy of the system. Equations (\ref{eq10}),
(\ref{eq11}), (\ref{eq13}), (\ref{eq16}), (\ref{eq17}) and
(\ref{eq19}) are fundamental correlation functions of the system.
When the right-hand sides of equations (\ref{eq8}), (\ref{eq9}),
(\ref{eq14}) and (\ref{eq15}) are expanded, the multispin
correlation functions can be easily obtained. The simplest
approximation, and one of the most frequently adopted is to decouple
these equations according to
\begin{equation}\label{eq24}
\left\langle
S_{i}^{z}(S_{j}^{z})^{2}...S_l^z\right\rangle\cong\left\langle
S_i^z\right\rangle\left\langle
(S_j^z)^{2}\right\rangle...\left\langle S_l^z\right\rangle,
\end{equation}
for $i\neq j \neq...\neq l$ \cite{decoupling}. The main difference
of the method used in this study from the other approximations in
the literature emerges in comparison with any decoupling
approximation (DA) when expanding the right-hand sides of equations
(\ref{eq8}), (\ref{eq9}), (\ref{eq14}) and (\ref{eq15}). In other
words, one advantage of the approximation method proposed by this
study is that no uncontrolled decoupling procedure is used for the
higher-order correlation functions.

For spin-1 Ising system with $q =3$, taking equations (\ref{eq10}),
(\ref{eq11}), (\ref{eq13}), (\ref{eq16}), (\ref{eq17}) and
(\ref{eq19}) as a basis, we derive a set of linear equations of the
spin correlation functions which interact in the system. At this
point, we assume that (i) the correlations depend only on the
distance between the spins, (ii) the average values of a central
spin and its nearest-neighbor spin (it is labeled as the perimeter
spin) are equal to each other, and (iii) in the matrix
representations of spin operator $\hat{S}$, the spin-1 system has
the properties $(S_{\delta}^{z})^{3}=S_{\delta}^{z}$ and
$(S_{\delta}^{z})^{4}=(S_{\delta}^{z})^{2}$. Thus, the number of the
set of linear equations obtained for the spin-1 Ising system with
$q=3$ reduces to twenty three and the complete set is given in
Appendix.

If equation (\ref{eqA7}) is written in the form of a $23\times23$
matrix and solved in terms of the variables
$x_{i}[(i=1,2,...,23)(e.g., x_{1}=\langle S_{0}^{z}\rangle,
x_{2}=\langle S_{1}S_{0}\rangle,..., x_{23}=\langle
(S_{0}^{x})^{2}\rangle)]$ of the linear equations, all of the spin
correlation functions can be easily determined as functions of the
temperature, effective field, crystal field and longitudinal
magnetic field as well as transverse magnetic field which the other
studies in the literature do not include. Since the thermal average
of the central spin is equal to that of its nearest-neighbor spins
within the present method then the unknown parameter $A$ can be
numerically determined by the relation
\begin{equation}\label{eq25}
\langle S_{0}^{z}\rangle=\langle S_{1}\rangle \qquad {\rm{ or
}}\qquad x_{1}=x_{4}.
\end{equation}

By solving equation (\ref{eq25}) numerically at a given fixed set of
Hamiltonian parameters we obtain the parameter $A$. Then we use the
numerical values of $A$ to obtain the spin correlation functions
$\langle S_{0}^{z}\rangle$, $\langle S_{0}^{x}\rangle$ (longitudinal
and transverse magnetizations), $\langle (S_{0}^{z})^{2}\rangle$,
$\langle (S_{0}^{x})^{2}\rangle$ (longitudinal and transverse
quadrupolar moments) and so on, which can be found from equation
(\ref{eqA7}). Note that $A=0$ is always the root of equation
(\ref{eq25}) corresponding to the disordered state of the system.
The nonzero root of $A$ in equation (\ref{eq25}) corresponds to the
long-range ordered state of the system. Once the spin correlation
functions have been evaluated then we can give the numerical results
for the thermal and magnetic properties of the system.

\section{Results and Discussions} \label{results}
\begin{figure}[]
  \subfigure[\hspace{0 cm}]
  {\includegraphics[width=6.5cm]{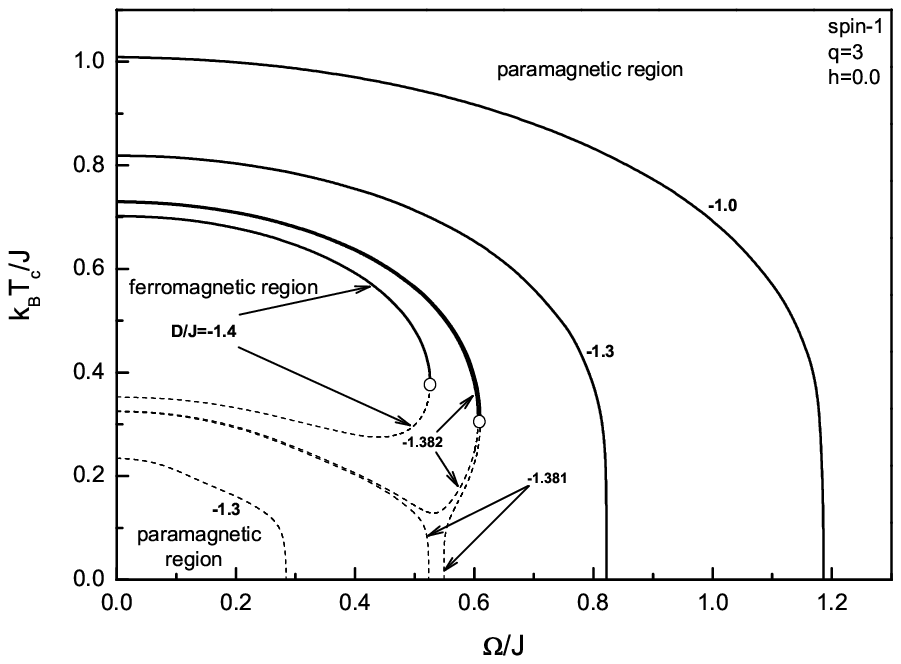}}\label{fig1a}
  \subfigure[\hspace{0 cm}] {\includegraphics[width=6.5cm]{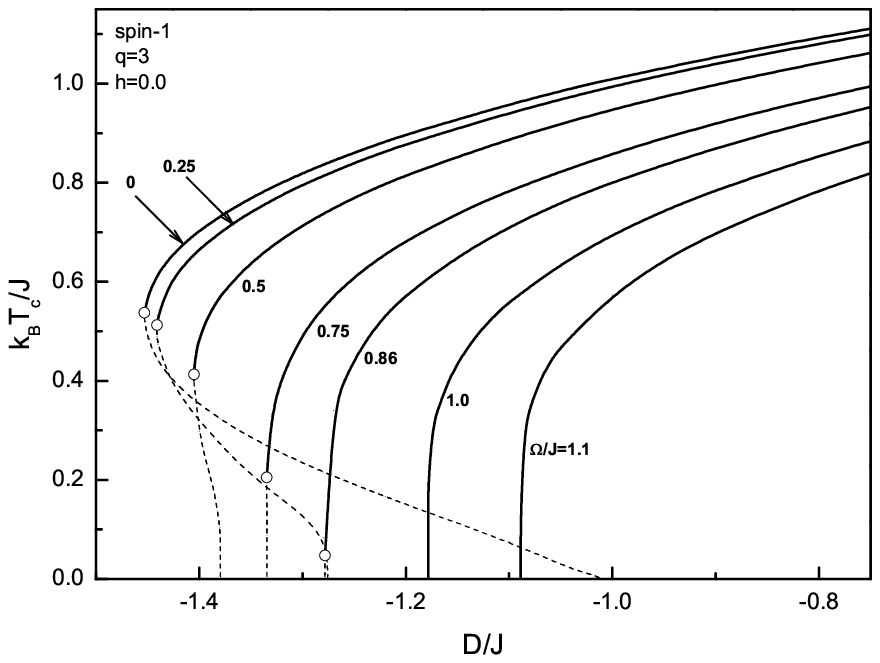}}\label{fig1b}\\
  \center
  \subfigure[\hspace{0 cm}] {\includegraphics[width=6.5cm]{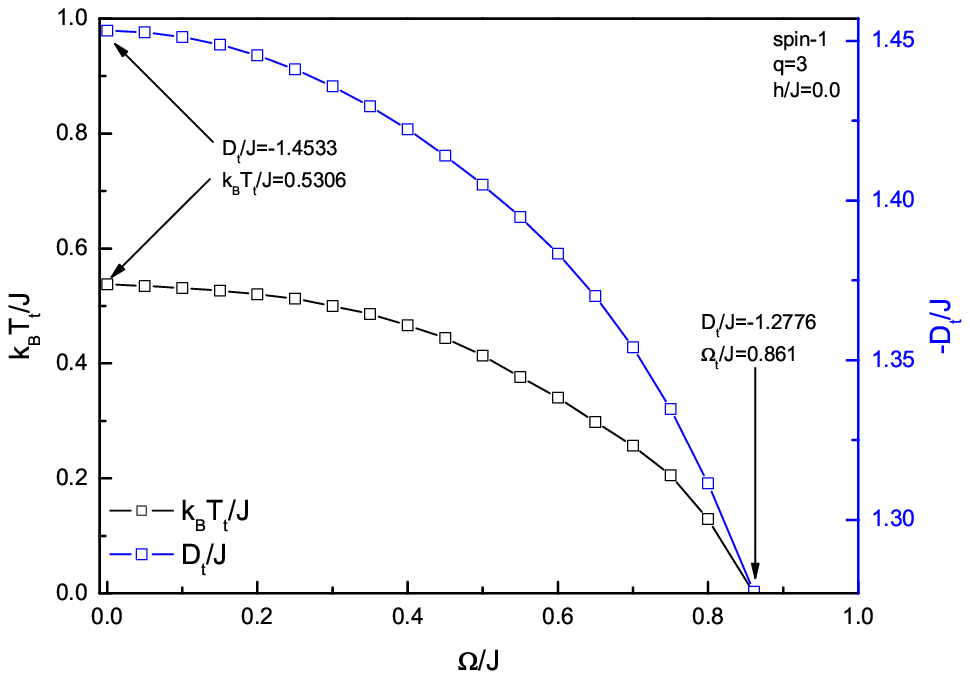}}\label{fig1c}\\
  \caption{ Phase diagrams of the spin-1 system with $h/J=0$ in (a) ($k_{B}T_{c}/J$-$\Omega/J$),
  (b) ($k_{B}T_{c}/J$-$D/J$) planes. The numbers on the curves denote the values of the crystal field
  $D/J$ and transverse field $\Omega/J$, respectively. (c) Transverse field dependencies of the tricritical
  temperature $k_{B}T_{t}/J$ and tricritical crystal field $-D_{t}/J$.  }
\end{figure}
In this section, we can examine the ferromagnetic properties of the
spin-1 TIM with crystal field under an applied longitudinal magnetic
field on a honeycomb lattice using the IEFT. For this purpose, we
focus our attention on the phase diagrams of the system in
$(k_{B}T_{c}/J-\Omega/J)$ and $(k_{B}T_{c}/J-D/J)$ planes and
investigate the whole phase diagrams by examining the numerical
results for the thermal and magnetic properties. In order to plot
the phase diagrams, we assume $\langle S_{0}^z\rangle=\langle
S_{1}\rangle$ and the effective field $\gamma$ is very small in the
vicinity of $k_{B}T_{c}/J$ and solve the set of linear equations in
equation (\ref{eqA7}) numerically using the self-consistent relation
corresponding to equation (\ref{eq25}). In Figs. 1a and 1b, we plot
the variation of the critical temperature with transverse field
$\Omega/J$ and crystal field $D/J$, respectively. Fig. 1a shows the
phase diagram in the $(k_{B}T_{c}/J-\Omega/J)$ plane with $h/J=0$
and for selected values of $D/J$, namely $-1.0, -1.3, -1.381,
-1.382$ and $-1.4$. In this figure, we can call attention to the
signs of an interesting behavior known as reentrant phenomena. In
other words, when the crystal field strength is positive valued, the
type of the transition in the system is invariably second order
which is independent from transverse field value. On the other hand,
if the crystal field value is sufficiently negative then we can
expect to see two successive phase transitions. Solid and dashed
lines in Fig. 1a correspond to the second and first order phase
transition lines, respectively. Tricritical end points at which
first and second order transition points meet are shown as white
circles. In our calculations, we realized that one can observe
reentrant behavior in the system for the values of $\Omega/J<0.861$
and $-1.4533<D/J<-1.0201$. For the values of $D/J\leq-1.382$ the
transition lines exhibit a bulge which gets smaller as the value of
$D/J$ approaches the value of $-1.4533$ which means that
ferromagnetic phase region gets narrower. We have also examined the
phase diagram of the present system in $(k_{B}T_{c}/J-D/J)$ plane
with $h/J=0$ and for selected values of $\Omega/J$ such as $0, 0.25,
0.5, 0.75, 0.86, 1.0$ and $1.1$. The numerical results are shown in
Fig. 1b. Solid and dashed lines in Fig. 1b correspond to the second
and first order phase transition lines, respectively. White circles
denote tricritical points. As we can see from this figure, as the
value of transverse field $\Omega/J$ increases starting from zero
then the value of tricritical point decreases gradually and
disappears for $\Omega/J>0.86$. If the transverse field value is
greater than this value then we have only second order transitions
in the system. These results show that the reentrant phenomenon
originates from the competition between the crystal field $D/J$ and
transverse field $\Omega/J$. Furthermore, the variation of the
coordinates of the tricritical points $k_{B}T_{t}/J$ and $D_{t}/J$
as a function of transverse field $\Omega/J$ is illustrated in Fig.
1c. This figure shows that the tricritical points exist for
$1.4333<-D_{t}/J<1.2776$ and $\Omega/J<0.861$. In addition, value of
$k_{B}T_{t}/J$ and the absolute value of $D_{t}/J$ decreases as the
value of transverse field increases and the tricritical temperature
disappears at the critical value of the transverse field
$\Omega_{t}/J=0.861$. All of the results mentioned above are in a
good agreement with other works
\cite{Saber,Jiang1,Htoutou2,Jiang2,Htoutou1,Htoutou3}, but not with
\cite{Miao}.

\begin{figure}[]
  \subfigure[\hspace{0 cm}] {\includegraphics[width=6.5cm]{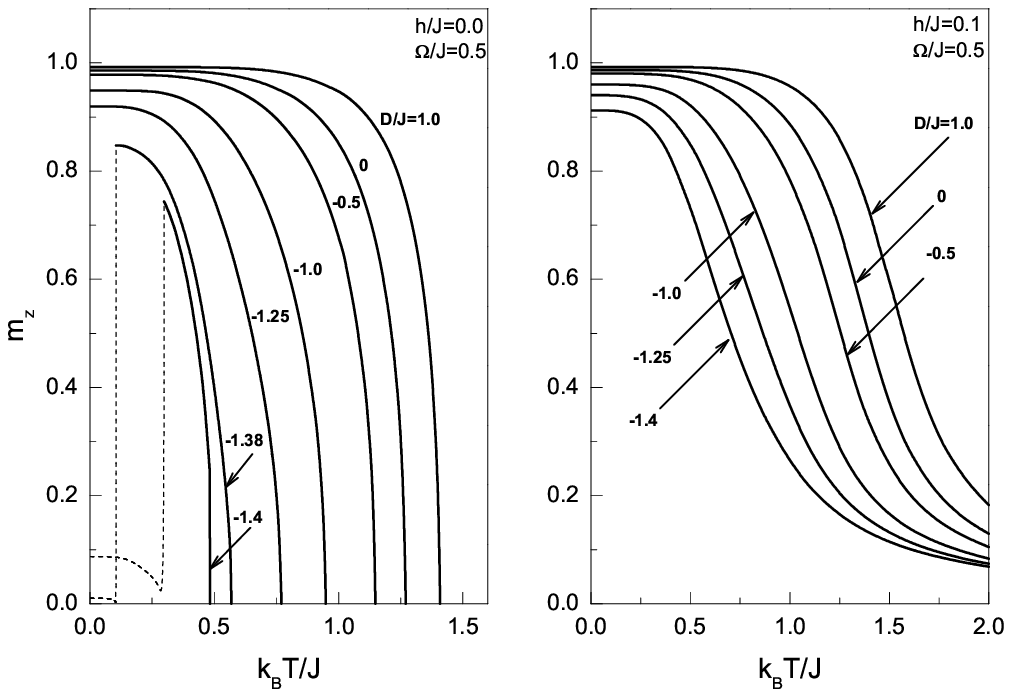}}
  \subfigure[\hspace{0 cm}] {\includegraphics[width=6.5cm]{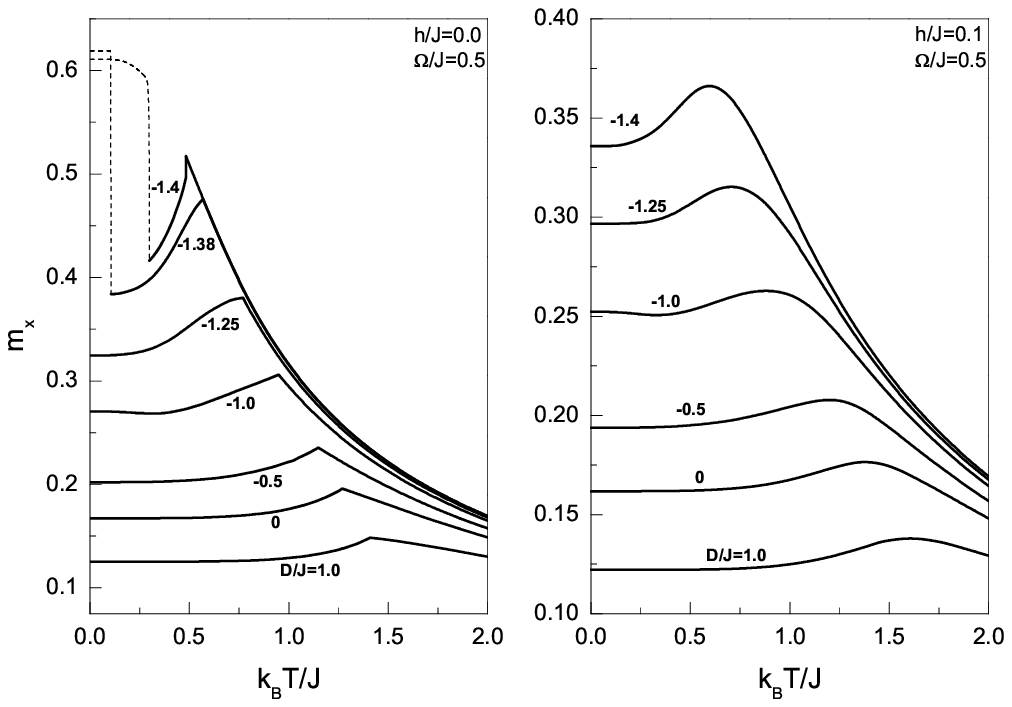}}\\
  \subfigure[\hspace{0 cm}] {\includegraphics[width=6.5cm]{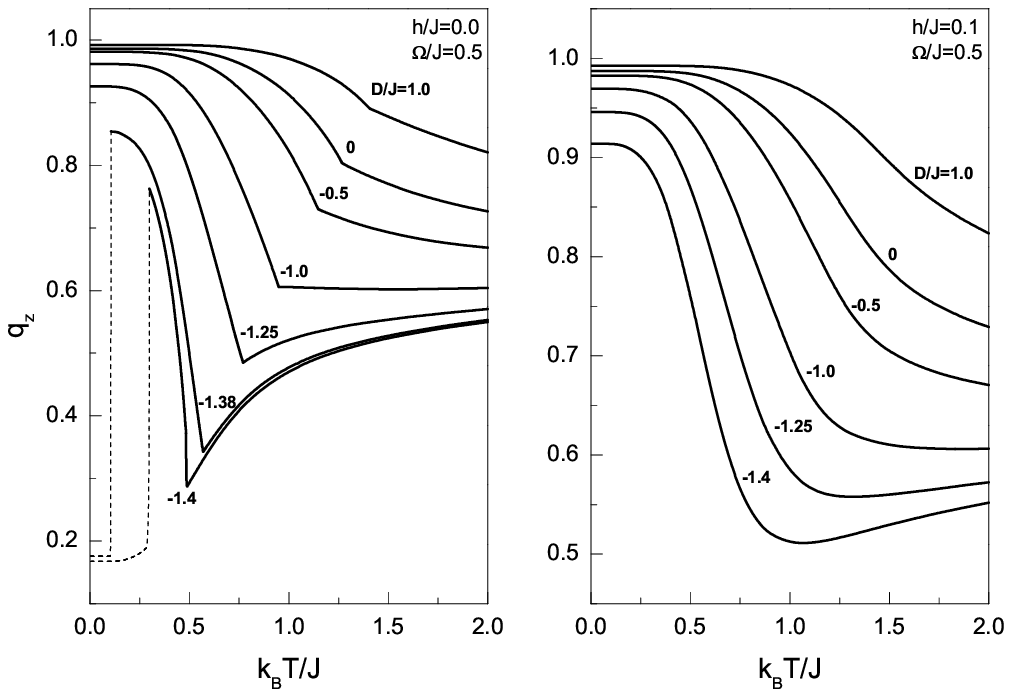}}
  \subfigure[\hspace{0 cm}] {\includegraphics[width=6.5cm]{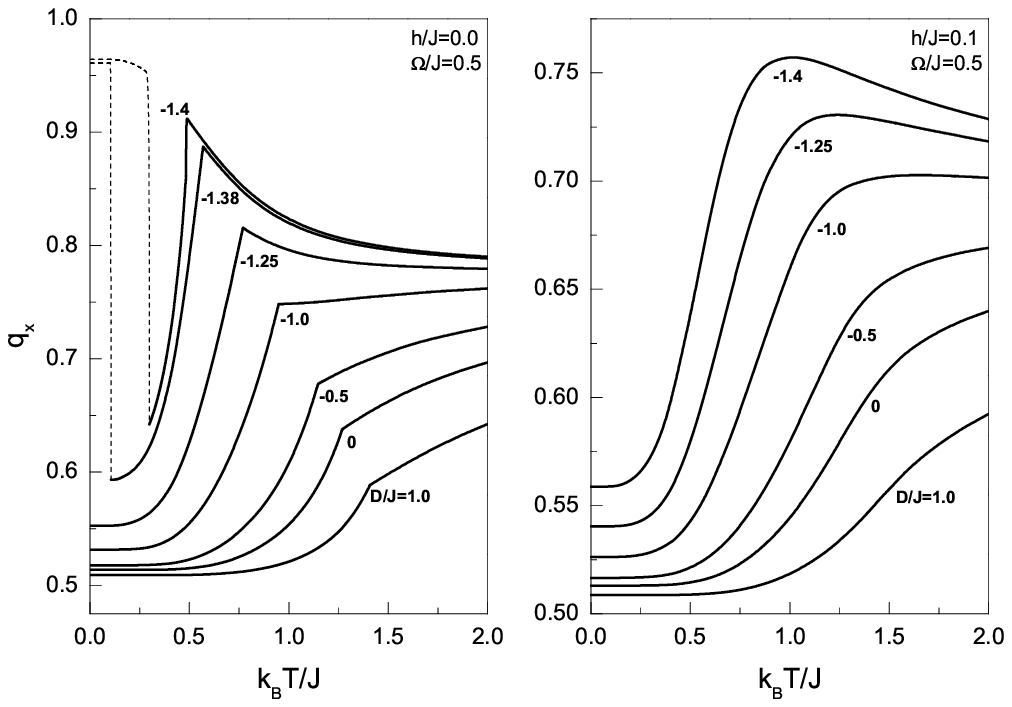}}\\
  \caption{ Temperature dependencies of (a) longitudinal magnetization
  $m_{z}$, (b) transverse magnetization $m_{x}$, (c) longitudinal quadrupolar moment
  $q_{z}$ and (d) transverse quadrupolar moment $q_{x}$ for the spin-1 system on a honeycomb
  lattice with a fixed value of transverse field $\Omega/J=0.5$ and some selected values of
  crystal field $D/J=1.0, 0, -0.5, -1.0, -1.25, -1.38$ and $-1.4$. The longitudinal magnetic field
  value is selected as $h/J=0$ and $0.1$ on the left and right panels, respectively. Dashed lines correspond to the first-order solutions.}
\end{figure}
In order to clarify the first-order phase transitions in the system
we examine the variation of the order parameters with temperature
such as the longitudinal and transverse magnetizations $m_{z}$ and
$m_{x}$ together with the longitudinal and transverse quadrupolar
moments $q_{z}$ and $q_{x}$. The effects of the crystal field on the
behavior of the order parameters with a selected transverse field
value $\Omega/J=0.5$ are shown in Fig. 2. Here, dashed lines
correspond to the solutions of the first order transition. As we can
see on the left panel in Fig. 2a, for the selected values of the
crystal field $D/J=1.0, 0, -0.5, -1.0$ and $-1.25$ with $h/J=0$ as
the temperature increases, the longitudinal magnetization $m_{z}$
falls rapidly from its saturation magnetization value at
$k_{B}T/J=0$ and decreases continuously in the vicinity of the
transition temperature and vanishes at a critical temperature
$T=T_{c}$. Besides, when we select the value of the crystal field
such as $D/J=-1.38$ and $-1.4$ we observe two successive phase
transitions. In other words, if we cool the system starting from a
finite temperature $T>T_{c}$, the system undergoes a phase
transition  from paramagnetic to ferromagnetic phase at $T=T_{c}$.
If we keep on cooling process then the second order transition at a
finite temperature is followed by a first order transition at a
lower temperature $T<T_{c}$. These results show the existence of
reentrant phenomena. Furthermore, as the value of crystal field
$D/J$ decreases then the second order transition temperature
decreases and the first order transition temperature increases. On
the left panel in Fig. 2b, we see the behavior of the transverse
magnetization with temperature for some selected values of crystal
field with $h/J=0$. For $D/J=1.0, 0, -0.5, -1.0$ and $-1.25$ the
transverse magnetization $m_{x}$ curves increase with the increase
of the temperature and then show a cusp which increases in height as
the value of the crystal field decreases at $T=T_{c}$ and decline as
the temperature increases. Consequently, the transverse
magnetization curves can be separated into two regions: the first is
the nonmagnetic region in which $m_{z}=0$ and the second is the
magnetic region in which $m_{z}\neq0$. For $D/J=-1.38$ and $-1.4$
$m_{x}$ curves show a discontinuous behavior at the first order
transition point. In Figs. 2c and 2d, the variation of the
longitudinal and transverse quadrupolar moments with temperature are
shown, respectively. The same crystal field $D/J$ values are used as
in Figs. 2a and 2b. We see that the longitudinal quadrupolar moment
$q_{z}$ decreases as the temperature increases and change abruptly
at the second order transition temperature for $D/J=1.0, 0, -0.5,
-1.0$ and $-1.25$. In case of $D/J=-1.38$ and $-1.4$, $q_{z}$ curves
exhibit two minima which correspond to the first and second order
transition temperatures. In contrast to $q_{z}$ curves in Fig. 2 c,
the transverse quadrupolar moment $q_{x}$ curves increases as the
temperature increases and change abruptly at the second order
transition temperature for $D/J=1.0, 0, -0.5, -1.0$ and $-1.25$. For
the values of crystal field $D/J=-1.38$ and $-1.4$ $q_{x}$ curves
exhibit two maxima corresponding to the first and second order
transitions in the system. When we apply a longitudinal magnetic
field such as $h/J=0.1$ on the system, both the first and second
order phase transition effects on the order parameters $m_{z}$,
$m_{x}$, $q_{z}$ and $q_{x}$ are removed. This phenomenon can be
clearly seen from the right hand side panels in Figs. 2a-2d.

\begin{figure}[]
  \subfigure[\hspace{0 cm}] {\includegraphics[width=6.5cm]{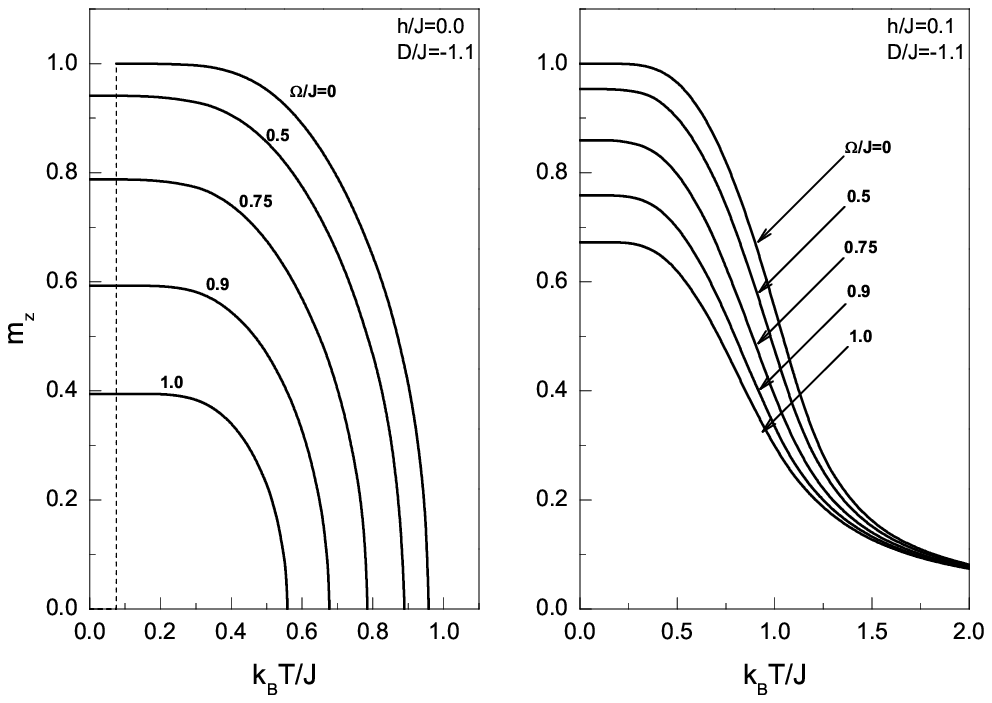}}
  \subfigure[\hspace{0 cm}] {\includegraphics[width=6.5cm]{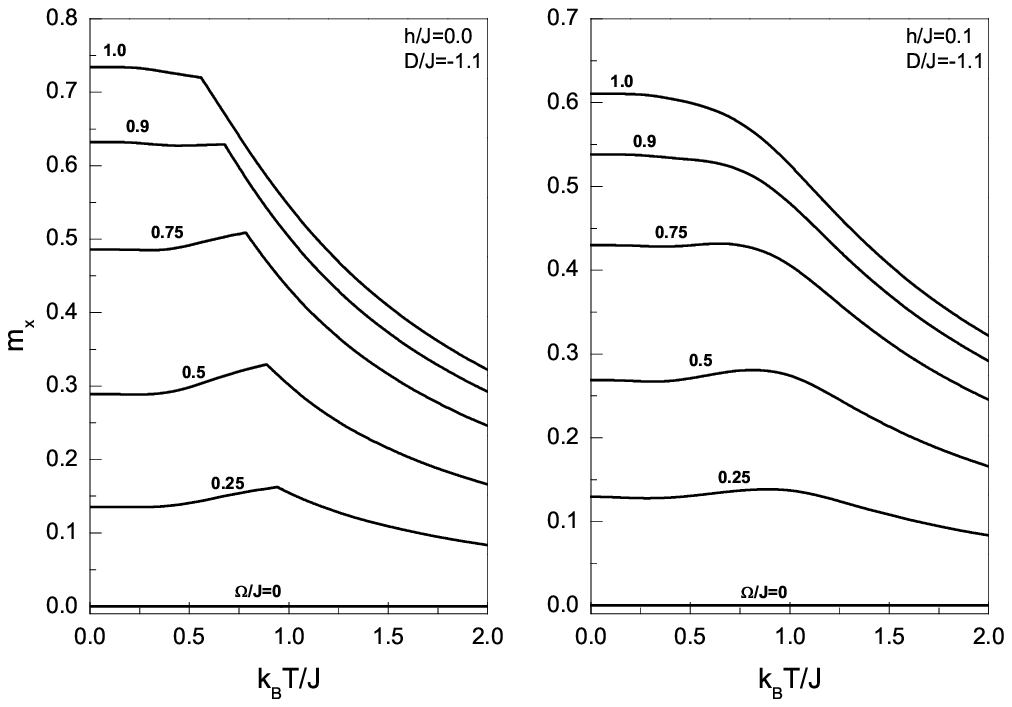}}\\
  \subfigure[\hspace{0 cm}] {\includegraphics[width=6.5cm]{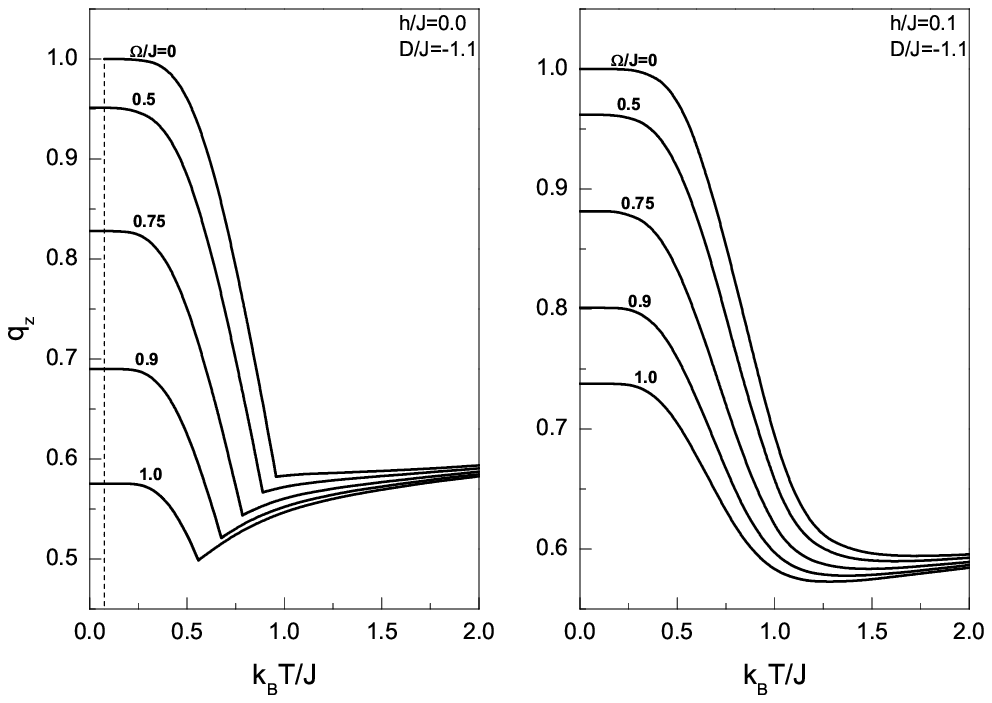}}
  \subfigure[\hspace{0 cm}] {\includegraphics[width=6.5cm]{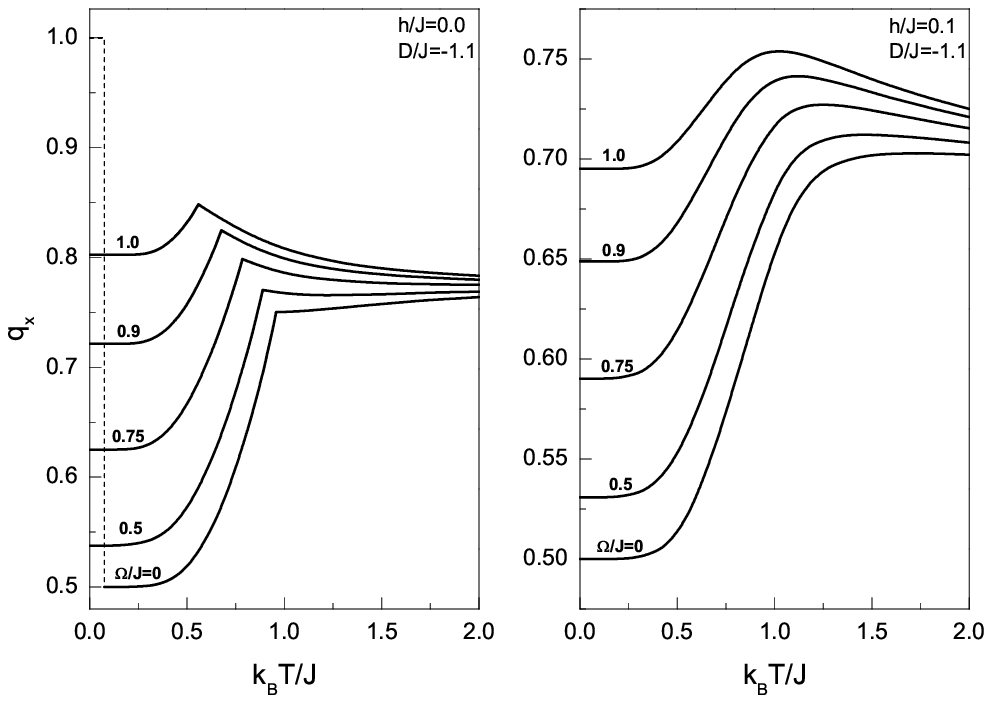}}\\
  \caption{Temperature dependencies of (a) longitudinal magnetization
  $m_{z}$, (b) transverse magnetization $m_{x}$, (c) longitudinal quadrupolar moment
  $q_{z}$ and (d) transverse quadrupolar moment $q_{x}$ for the spin-1 system on a honeycomb
  lattice with a fixed value of crystal field $D/J=-1.1$ and some selected values of
  transverse field $\Omega/J=0, 0.5, 0.75, 0.9$ and $1.0$. $\Omega/J=0.25$ is also selected in (b). The longitudinal magnetic field
  value is selected as $h/J=0$ and $0.1$ on the left and right panels, respectively. Dashed lines correspond to the first-order solutions.}
\end{figure}
Next, the transverse field effect on the variation of the order
parameters with temperature such as the longitudinal and transverse
magnetizations $m_{z}$ and $m_{x}$ as well as the longitudinal and
transverse quadrupolar moments $q_{z}$ and $q_{x}$ with $D/J=-1.1$
and $h/J=0$ can be seen on the left panels Fig. 3. Dashed lines
represent the first order transition solutions. In Fig. 3a, we plot
the longitudinal component of magnetization for selected values of
transverse field $\Omega/J=0, 0.5, 0.75, 0.9$ and $1$. As we can see
on the left panel in Fig. 3a, as the transverse field $\Omega/J$
increases, critical temperature value approaches to zero and
saturation value of $m_{z}$ decreases. Hence, we can say that
applying transverse field $\Omega/J$ weakens the longitudinal
component of magnetization. This is an expected result. However, two
successive transitions (i.e. reentrant behavior)  are observed on
the longitudinal magnetization $m_{z}$ for $\Omega/J=0$. It is clear
that applied transverse field $\Omega/J$ destructs the first order
transition. On the left panel in Fig. 3b, we represent the effect of
the transverse field $\Omega/J$ on the temperature dependence of
$m_{x}$. In contrast to the situation in $m_{z}$, transverse field
$\Omega/J$ strengthens the transverse component of magnetization. We
note that, at zero transverse field the transverse magnetization
$m_{x}$ is zero for the whole range of temperature. On the left
panels in Figs. 3c and 3d, we present the effect of the transverse
field $\Omega/J$ on the variation of the quadrupolar moments $q_{z}$
and $q_{x}$ with temperature, respectively. As we can clearly see
from these plots, longitudinal quadrupolar moment $q_{z}$ has two
minima while transverse counterpart $q_{x}$ has two maxima
corresponding to the first and second order transitions for
$\Omega/J=0$. Furthermore, applying any longitudinal magnetic field
such as $h/J=0.1$ on the system, both the first and second order
phase transitions are destructed. This behavior is illustrated on
the right panels in Fig. 3. Hence, on the basis of these results
(see the right panels in Figs. 2 and 3) we believe that the effects
of the transverse field $\Omega/J$ are very different from those of
the longitudinal counterpart $h/J$ because the origin of the
transverse field is quantum mechanical and can produce quantum
effects.

\begin{figure}
  \subfigure[\hspace{0 cm}] {\includegraphics[width=6.5cm]{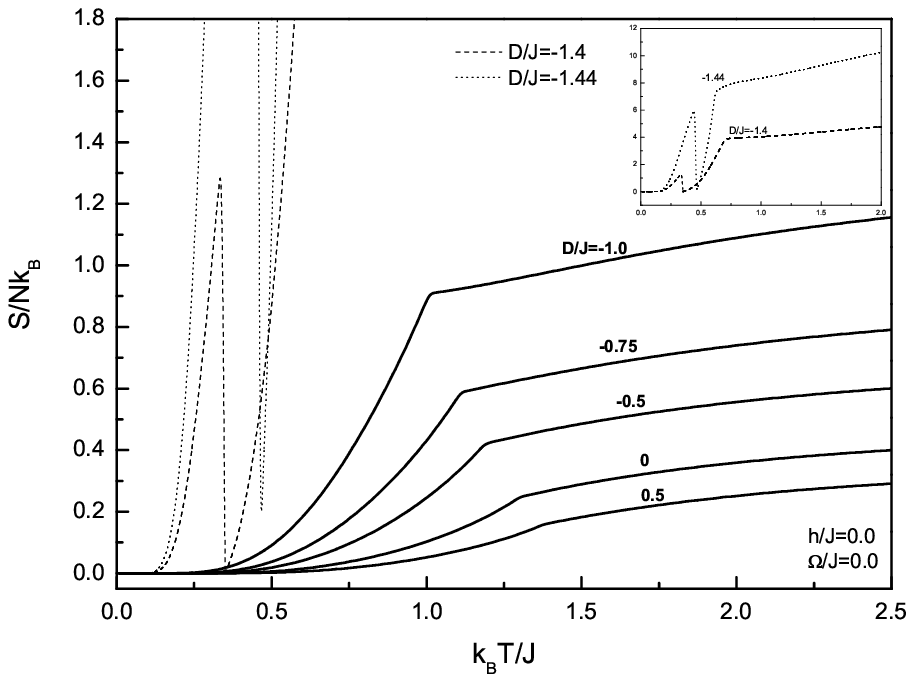}}
  \subfigure[\hspace{0 cm}] {\includegraphics[width=6.5cm]{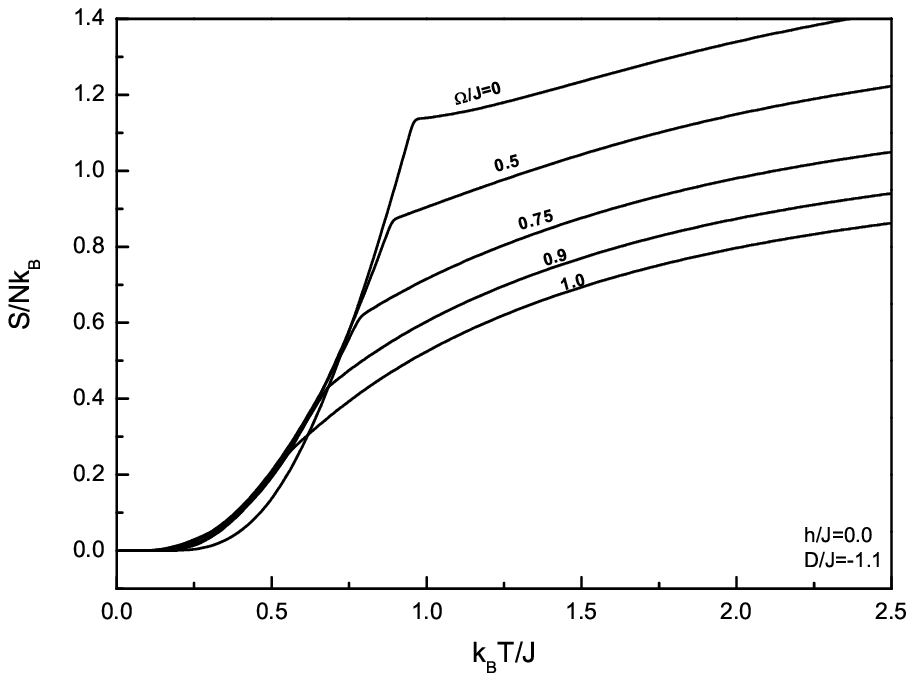}}\\
  \subfigure[\hspace{0 cm}] {\includegraphics[width=6.5cm]{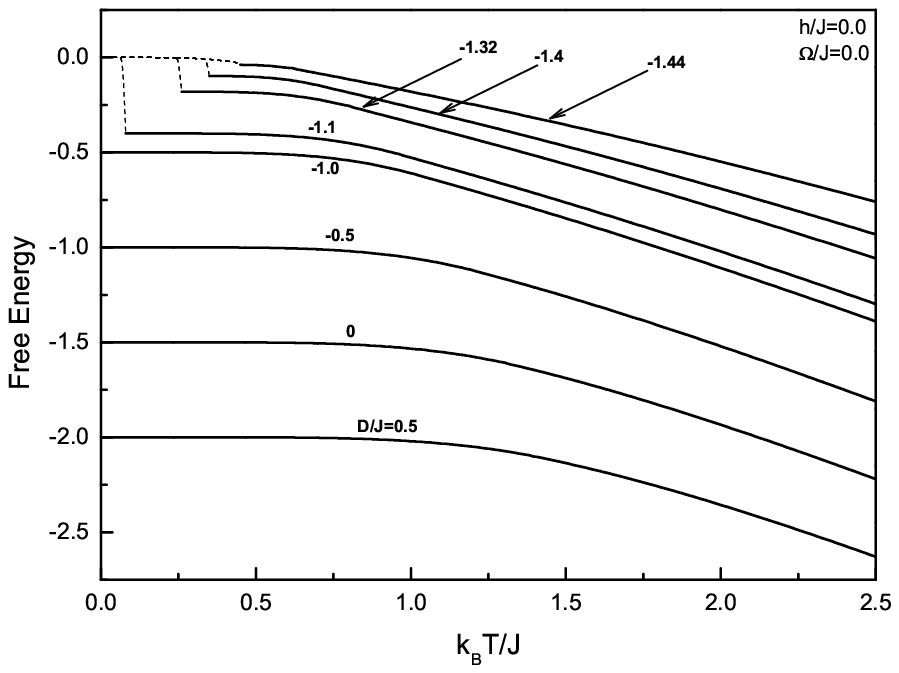}}
  \subfigure[\hspace{0 cm}] {\includegraphics[width=6.5cm]{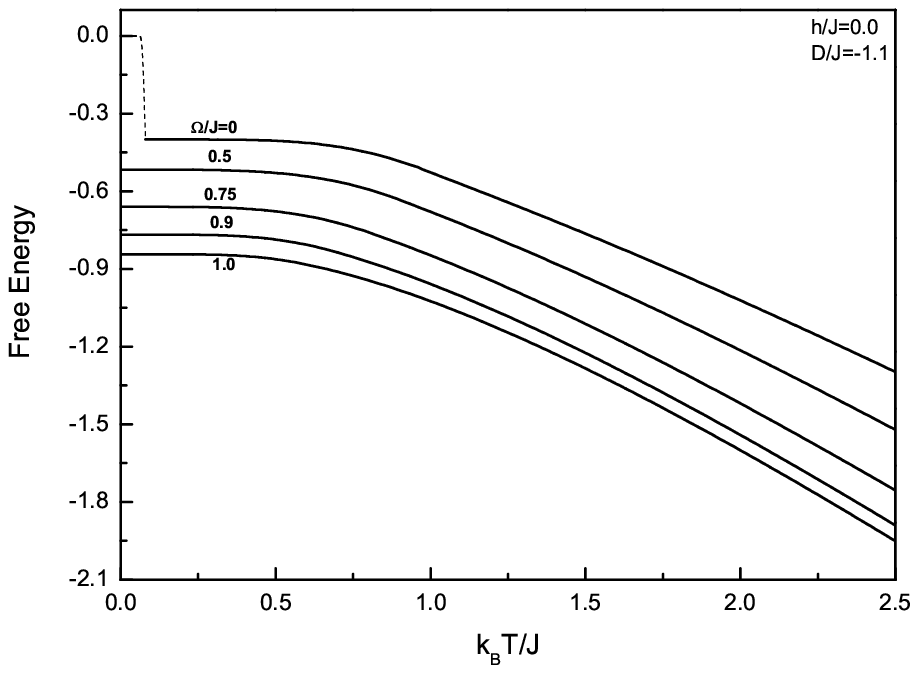}}\\
  \caption{ Entropy of the system as a function of temperature for $h/J=0$ and (a) $\Omega/J=0$
  with some selected values of crystal field $D/J=0.5, 0, -0.5, -0.75, -1.0, -1.4$
  and $-1.44$, (b) $\Omega/J=0, 0.5, 0.75, 0.9$ and $1.0$ with a fixed value of crystal field
  $D/J=-1.1$. (c) Temperature dependence of Helmholtz free energy of the system for $h/J=0$.
  The value of transverse field is fixed as $\Omega/J=0.0$ with some
  selected values of crystal field $D/J=0.5, 0, -0.5, -1.0, -1.1, -1.32, -1.4$ and
  $-1.44$. (d) crystal field is fixed as $D/J=-1.1$ with five different
  values of transverse field $\Omega/J=0, 0.5, 0.75, 0.9$ and $1.0$. }
\end{figure}
Finally in Fig. 4, we present the numerical results for the
temperature dependence of entropy per site and Helmholtz free energy
of the system. As far as we know, there is not such a study dealing
with the variation of the entropy of the system with the Hamiltonian
parameters. In Figs. 4a and 4c, we show the effect of the crystal
field $D/J$ on the entropy and free energy for $\Omega/J=0$ and
$h/J=0$. For $D/J=0.5, 0, -0.5, -0.75$ and $-1.0$ entropy is not
important at low temperatures and free energy is equal to ground
state energy of the system. However, as the temperature increases,
the system wants to maximize its entropy in order to minimize its
free energy and hence entropy becomes important. Furthermore, the
entropy of the system is continuous at the critical temperature
which means that the type of the transition is second order. On the
other hand, for $D/J<-1.0$ the entropy and free energy curves
represented with dashed lines show a discontinuous behavior at low
temperatures which indicates a first order transition and it
originates from a discontinuous change in the internal energy of the
system. Discontinuous behavior of the entropy can be clearly seen in
the inset figure in Fig. 4a. As the absolute value of the crystal
field increases, energy gap in free energy curves (see Fig. 4c)
disappear and ground state energy equals to zero for tricritical
crystal field value $D_{t}/J$. Furthermore, we see that increasing
the absolute value of the crystal field $D/J$ makes the absolute
value of free energy decrease and causes the system to be in a
considerably disordered state. These remarkable observations are not
reported in the literature. The effect of the transverse field
$\Omega/J$ on the temperature dependence of entropy and Helmholtz
free energy can be seen in Figs. 4b and 4d for selected values of
$\Omega/J=0, 0.5, 0.75, 0.9$ and $1.0$ with $D/J=-1.1$ and $h/J=0$.
As the transverse field $\Omega/J$ value increases, critical
temperature decreases and the system reaches disordered phase and
maximizes its entropy while minimizing its free energy at early
stages of temperature scale. Besides, applied transverse field
$\Omega/J$ tends to destroy the energy gap in free energy and hence
the first order transitions are removed. One should notice that
although we observe the first order transition effects for
$\Omega/J=0$ in free energy  (see Fig. 4c), this effect is not
evident for the entropy (Fig. 4b). This is due to the fact that,
when we select $D/J=-1.1$ and $h/J=0$ with $\Omega/J=0$, the system
is in a disordered state and the internal energy is zero at low
temperatures (i.e. ground state energy) and its value is not changed
until the system undergoes a first order transition.

\section{Conclusion} \label{conclusion}
In this work, we have studied the phase diagrams of the spin-1
transverse Ising model with a longitudinal crystal field in the
presence of a longitudinal magnetic field on a honeycomb lattice
within the framework of the IEFT approximation that takes into
account the correlations between different spins in the cluster of
considered lattice. We have given the proper phase diagrams,
especially the first-order transition lines that include reentrant
phase transition regions. Both the order parameters ($m_{z}$,
$m_{x}$, $q_{z}$, $q_{x}$) and Helmholtz free energy $F$ and the
entropy $S$ curves show discontinuous and unstable points of the
system and support the predictions in Figs. 1a and 1b.

A number of interesting phenomena such as reentrant phenomena have
been found in the physical quantities originating from the crystal
field as well as the transverse and longitudinal components of the
magnetic field. We have found that one can observe reentrant
behavior in the system for the values of $\Omega/J<0.861$ and
$-1.4533<D/J<-1.0201$ and the tricritical points exist for
$1.4333<-D_{t}/J<1.2776$ and $\Omega/J<0.861$. The results show that
the reentrant phenomenon originates from the competition between the
crystal field $D/J$ and transverse field $\Omega/J$. Besides,
applying a transverse field $\Omega/J$ on the system has the
tendency to destruct the first order transitions, while the
longitudinal counterpart $h/J$ destructs both the first and second
order phase transitions. Hence, we believe that the effects of the
transverse field $\Omega/J$ are very different from those of the
longitudinal counterpart $h/J$ since the transverse field can
produce quantum effects. These interesting results are not reported
in the literature.

We hope that the results obtained in this work may be beneficial
from both theoretical and experimental point of view.

\section*{Acknowledgements}
One of the authors (YY) would like to thank the Scientific and
Technological Research Council of Turkey (T\"{U}B\.{I}TAK) for
partial financial support. This work has been completed at the Dokuz
Eyl\"{u}l University, Graduate School of Natural and Applied
Sciences and is the subject of the forthcoming Ph.D. thesis of Y.
Y\"{u}ksel.

\newpage

\appendix
\label{appendix}
\section{}
The basis relations corresponding to equations (\ref{eq8}),
(\ref{eq9}), (\ref{eq12}) (\ref{eq14}) (\ref{eq15}) and
(\ref{eq18}). The coefficients $l_{i}$, $p_{i}$, $r_{i}$, $\nu_{i}$
($i=0,1,...,5$); $k_{j}$, $c_{j}$, $n_{j}$, $\mu_{j}$ ($j=1,...,4$);
and $a_{k}$, $b_{k}$ ($k=1,2,3$) can be derived from a mathematical
identity $\exp(\alpha\nabla)F(x)=F(x+\alpha)$.

\begin{eqnarray}\label{eqA1}
\nonumber \langle S_{0}^{z}\rangle&=&l_{0}+3k_{1}\langle S_{1}
\rangle+3(l_{1}-l_{0})\langle S_{1}^{2} \rangle+3l_{2}\langle
S_{1}S_{2}
\rangle \\
\nonumber& & +6(k_{2}-k_{1})\langle S_{1}S_{2}^{2}
\rangle+3(l_{0}-2l_{1}+l_{3})\langle S_{1}^{2}S_{2}^{2}
\rangle \\
\nonumber& &+k_{3}\langle S_{1}S_{2}S_{3}
\rangle+3(l_{4}-l_{2})\langle S_{1}S_{2}S_{3}^{2}
\rangle\\
\nonumber& &+3(k_{1}-2k_{2}+k_{4})\langle S_{1}S_{2}^{2}S_{3}^{2}
\rangle\\
& &+(-l_{0}+3l_{1}-3l_{3}+l_{5})\langle S_{1}^{2}S_{2}^{2}S_{3}^{2}
\rangle
\end{eqnarray}
\begin{eqnarray}
\nonumber  l_{0}&=&F(0)  \\
\nonumber  l_{1}&=& \mathrm{cosh}(J\nabla)F(x)|_{x=0} \\
\nonumber  l_{2}&=& \mathrm{sinh}^{2}(J\nabla)F(x)|_{x=0} \\
\nonumber  l_{3}&=& \mathrm{cosh}^{2}(J\nabla)F(x)|_{x=0} \\
\nonumber  l_{4}&=& \mathrm{cosh}(J\nabla) \mathrm{sinh}^{2}(J\nabla)F(x)|_{x=0} \\
\nonumber  l_{5}&=&\mathrm{cosh}^{3}(J\nabla)F(x)|_{x=0}
\end{eqnarray}
\begin{eqnarray}
\nonumber k_{1}&=& \mathrm{sinh}(J\nabla)F(x)|_{x=0}\\
\nonumber k_{2}&=& \mathrm{cosh}(J\nabla)
\mathrm{sinh}(J\nabla)F(x)|_{x=0}\\
\nonumber k_{3}&=& \mathrm{sinh}^{3}(J\nabla)F(x)|_{x=0}\\
\nonumber k_{4}&=& \mathrm{cosh}^{2}(J\nabla)
\mathrm{sinh}(J\nabla)F(x)|_{x=0}
\end{eqnarray}

\begin{eqnarray}\label{eqA2}
\nonumber \langle S_{0}^{x}\rangle&=&p_{0}+3c_{1}\langle S_{1}
\rangle+3(p_{1}-p_{0})\langle S_{1}^{2} \rangle+3p_{2}\langle
S_{1}S_{2}
\rangle \\
\nonumber& & +6(c_{2}-c_{1})\langle S_{1}S_{2}^{2}
\rangle+3(p_{0}-2p_{1}+p_{3})\langle S_{1}^{2}S_{2}^{2}
\rangle \\
\nonumber& &+c_{3}\langle S_{1}S_{2}S_{3}
\rangle+3(p_{4}-p_{2})\langle S_{1}S_{2}S_{3}^{2}
\rangle\\
\nonumber& &+3(c_{1}-2c_{2}+c_{4})\langle S_{1}S_{2}^{2}S_{3}^{2}
\rangle\\
& &+(-p_{0}+3p_{1}-3p_{3}+p_{5})\langle S_{1}^{2}S_{2}^{2}S_{3}^{2}
\rangle
\end{eqnarray}
\begin{eqnarray}
\nonumber  p_{0}&=&H(0)\\
\nonumber  p_{1}&=& \mathrm{cosh}(J\nabla)H(x)|_{x=0}  \\
\nonumber  p_{2}&=& \mathrm{sinh}^{2}(J\nabla)H(x)|_{x=0} \\
\nonumber  p_{3}&=& \mathrm{cosh}^{2}(J\nabla)H(x)|_{x=0}\\
\nonumber  p_{4}&=& \mathrm{cosh}(J\nabla) \mathrm{sinh}^{2}(J\nabla)H(x)|_{x=0} \\
\nonumber  p_{5}&=& \mathrm{cosh}^{3}(J\nabla)H(x)|_{x=0}
\end{eqnarray}
\begin{eqnarray}
\nonumber c_{1}&=& \mathrm{sinh}(J\nabla)H(x)|_{x=0}\\
\nonumber c_{2}&=& \mathrm{cosh}(J\nabla) \mathrm{sinh}(J\nabla)H(x)|_{x=0}\\
\nonumber c_{3}&=& \mathrm{sinh}^{3}(J\nabla)H(x)|_{x=0}\\
\nonumber c_{4}&=& \mathrm{cosh}^{2}(J\nabla)
\mathrm{sinh}(J\nabla)H(x)|_{x=0}
\end{eqnarray}

\begin{equation}\label{eqA3}
\langle S_{1}\rangle=a_{1}\left(1-\langle
(S_{0}^{z})^{2}\rangle\right)+a_{2}\langle S_{0}^{z}
\rangle+a_{3}\langle (S_{0}^{z})^{2}\rangle
\end{equation}
\begin{eqnarray}
\nonumber  a_{1} &=& F(\gamma) \\
\nonumber  a_{2} &=&  \mathrm{sinh}(J\nabla)F(x+\gamma)|_{x=0} \\
\nonumber  a_{3} &=& \mathrm{cosh}(J\nabla)F(x+\gamma)|_{x=0}
\end{eqnarray}

\begin{eqnarray}\label{eqA4}
\nonumber  \langle (S_{0}^{z})^{2}\rangle &=& r_{0}+3n_{1}\langle S_{1}\rangle+3(r_{1}-r_{0})\langle S_{1}^{2}\rangle+3r_{2}\langle S_{1}S_{2}\rangle \\
\nonumber   &&+6(n_{2}-n_{1})\langle S_{1}S_{2}^{2}\rangle+3(r_{0}-2r_{1}+r_{3})\langle S_{1}^{2}S_{2}^{2}\rangle \\
\nonumber   &&+n_{3}\langle S_{1}S_{2}S_{3}\rangle+ 3(r_{4}-r_{2})\langle S_{1}S_{2}S_{3}^{2}\rangle  \\
\nonumber   &&+3(n_{1}-2n_{2}+n_{4})\langle
S_{1}S_{2}^{2}S_{3}^{2}\rangle\\
   &&+(-r_{0}+3r_{1}-3r_{3}+r_{5})\langle
   S_{1}^{2}S_{2}^{2}S_{3}^{2}\rangle
\end{eqnarray}
\begin{eqnarray}
\nonumber  r_{0}&=& G(0) \\
\nonumber  r_{1}&=& \mathrm{cosh}(J\nabla)G(x)|_{x=0} \\
\nonumber  r_{2}&=& \mathrm{sinh}^{2}(J\nabla)G(x)|_{x=0} \\
\nonumber  r_{3}&=& \mathrm{cosh}^{2}(J\nabla)G(x)|_{x=0} \\
\nonumber  r_{4}&=& \mathrm{cosh}(J\nabla) \mathrm{sinh}^{2}(J\nabla)G(x)|_{x=0} \\
\nonumber  r_{5}&=& \mathrm{cosh}^{3}(J\nabla)G(x)|_{x=0}
\end{eqnarray}

\begin{eqnarray}
\nonumber n_{1}&=& \mathrm{sinh}(J\nabla)G(x)|_{x=0}\\
\nonumber n_{2}&=& \mathrm{cosh}(J\nabla) \mathrm{sinh}(J\nabla)G(x)|_{x=0}\\
\nonumber n_{3}&=& \mathrm{sinh}^{3}(J\nabla)G(x)|_{x=0} \\
\nonumber n_{4}&=& \mathrm{cosh}^{2}(J\nabla)
\mathrm{sinh}(J\nabla)G(x)|_{x=0}
\end{eqnarray}

\begin{eqnarray}\label{eqA5}
\nonumber \langle (S_{0}^{x})^2\rangle&=&v_{0}+3\mu_{1}\langle S_{1}
\rangle+3(v_{1}-v_{0})\langle S_{1}^{2} \rangle+3v_{2}\langle
S_{1}S_{2}
\rangle \\
\nonumber& & +6(\mu_{2}-\mu_{1})\langle S_{1}S_{2}^{2}
\rangle+3(v_{0}-2v_{1}+v_{3})\langle S_{1}^{2}S_{2}^{2}
\rangle \\
\nonumber& &+\mu_{3}\langle S_{1}S_{2}S_{3}
\rangle+3(v_{4}-v_{2})\langle S_{1}S_{2}S_{3}^{2}
\rangle\\
\nonumber& &+3(\mu_{1}-2\mu_{2}+\mu_{4})\langle
S_{1}S_{2}^{2}S_{3}^{2}
\rangle\\
& &+(-v_{0}+3v_{1}-3v_{3}+v_{5})\langle S_{1}^{2}S_{2}^{2}S_{3}^{2}
\rangle
\end{eqnarray}
\begin{eqnarray}
\nonumber  v_{0}&=&K(0)\\
\nonumber  v_{1}&=& \mathrm{cosh}(J\nabla)K(x)|_{x=0} \\
\nonumber  v_{2}&=& \mathrm{sinh}^{2}(J\nabla)K(x)|_{x=0} \\
\nonumber  v_{3}&=& \mathrm{cosh}^{2}(J\nabla)K(x)|_{x=0}  \\
\nonumber  v_{4}&=& \mathrm{cosh}(J\nabla) \mathrm{sinh}^{2}(J\nabla)K(x)|_{x=0}  \\
\nonumber  v_{5}&=& \mathrm{cosh}^{3}(J\nabla)K(x)|_{x=0}
\end{eqnarray}
\begin{eqnarray}
\nonumber \mu_{1}&=& \mathrm{sinh}(J\nabla)K(x)|_{x=0} \\
\nonumber \mu_{2}&=& \mathrm{cosh}(J\nabla) \mathrm{sinh}(J\nabla)K(x)|_{x=0}\\
\nonumber\mu_{3}&=& \mathrm{sinh}^{3}(J\nabla)K(x)|_{x=0}\\
\nonumber\mu_{4}&=& \mathrm{cosh}^{2}(J\nabla)
\mathrm{sinh}(J\nabla)K(x)|_{x=0}
\end{eqnarray}

\begin{equation}\label{eqA6}
\langle S_{1}^{2}\rangle=b_{1}\left(1-\langle
(S_{0}^{z})^{2}\rangle\right)+b_{2}\langle S_{0}^{z}
\rangle+b_{3}\langle (S_{0}^{z})^{2}\rangle
\end{equation}
\begin{eqnarray}
\nonumber  b_{1} &=& G(\gamma) \\
\nonumber  b_{2} &=& \mathrm{sinh}(J\nabla)G(x+\gamma)|_{x=0} \\
\nonumber  b_{3} &=& \mathrm{cosh}(J\nabla)G(x+\gamma)|_{x=0}
\end{eqnarray}
The complete set of twenty three linear equations of the spin-1
honeycomb lattice $(q=3)$:
\begin{eqnarray}
\nonumber\langle S_{0}^{z}\rangle&=&l_{0}+3k_{1}\langle S_{1}
\rangle+3(l_{1}-l_{0})\langle S_{1}^{2} \rangle+3l_{2}\langle
S_{1}S_{2}
\rangle \\
\nonumber& & +6(k_{2}-k_{1})\langle S_{1}S_{2}^{2}
\rangle+3(l_{0}-2l_{1}+l_{3})\langle S_{1}^{2}S_{2}^{2}
\rangle \\
\nonumber& &+k_{3}\langle S_{1}S_{2}S_{3}
\rangle+3(l_{4}-l_{2})\langle S_{1}S_{2}S_{3}^{2}
\rangle\\
\nonumber& &+3(k_{1}-2k_{2}+k_{4})\langle S_{1}S_{2}^{2}S_{3}^{2}
\rangle\\
\nonumber& &+(-l_{0}+3l_{1}-3l_{3}+l_{5})\langle
S_{1}^{2}S_{2}^{2}S_{3}^{2} \rangle\\
\nonumber \langle S_{1}S_{0}\rangle &=& (3l_{1}-2l_{0})\langle
S_{1}\rangle+3k_{1}\langle
S_{1}^{2}\rangle \\
\nonumber&&+3(l_{0}-2l_{1}+l_{2}+l_{3})\langle S_{1}S_{2}^{2}\rangle\\
\nonumber&&+6(k_{2}-k_{1})\langle
S_{1}^2S_{2}^{2}\rangle+k_{3}\langle
S_{1}S_{2}S_{3}^{2}\rangle\\
\nonumber&&+(-l_{0}+3l_{1}-3l_{2}-3l_{3}+3l_{4}+l_{5})\langle
S_{1}S_{2}^{2}S_{3}^{2}\rangle\\
\nonumber&&+3(k_{1}-2k_{2}+k_{4})\langle
S_{1}^{2}S_{2}^{2}S_{3}^{2}\rangle\\
\nonumber \langle
S_{1}S_{2}S_{0}\rangle &=& ({l_0}-3{l_1}+3{l_2}+3{l_3})\langle{S_1}{S_2}\rangle+(6{k_2}-3{k_1})\langle{S_1}S_{2}^{2}\rangle \\
\nonumber&&+
(-{l_0}+3{l_1}-3{l_2}-3{l_3}+3{l_4}+{l_5})\langle{S_1}{S_2}S_{3}^{2}\rangle\\
\nonumber&&+(3{k_1}-6{k_2}+{k_3}+3{k_4})\langle{S_1}S_{2}^{2}S_{3}^{2}\rangle\\
\nonumber \langle S_{1}\rangle &=& a_{1}(1-\langle
(S_{0}^{z})^{2}\rangle)+a_{2}\langle S_{0}^{z} \rangle+a_{3}\langle
(S_{0}^{z})^{2}\rangle \\
\nonumber  \langle S_{1}S_{2}\rangle &=& a_{1} \langle S_{1}
\rangle+a_{2}\langle S_{0}S_{1}\rangle+(a_{3}-a_{1})\langle S_{1}S_{0}^{2}\rangle \\
\nonumber  \langle S_{1}S_{2}S_{3}\rangle &=& a_{1}\langle
S_{1}S_{2}\rangle
+a_{2}\langle S_{0}S_{1}S_{2}\rangle+(a_{3}-a_{1})\langle S_{1}S_{2}S_{0}^{2}\rangle  \\
\nonumber \langle S_{1}^{2}\rangle&=&b_{1}(1-\langle
(S_{0}^{z})^{2}\rangle)+b_{2}\langle S_{0}^{z} \rangle+b_{3}\langle
(S_{0}^{z})^{2}\rangle\\
\nonumber  \langle S_{1}S_{2}^2\rangle &=& b_{1}\langle S_{1}
\rangle+b_{2}\langle S_{0}S_{1}\rangle+(b_{3}-b_{1}) \langle S_{1}S_{0}^2\rangle  \\
\nonumber  \langle S_{1}^2S_{2}^2\rangle &=& b_{1}\langle
S_{1}^2\rangle+b_{2}
\langle S_{0}S_{1}^2\rangle+(b_{3}-b_{1})\langle S_{1}^2S_{0}^2\rangle  \\
\nonumber  \langle S_{0}S_{1}^2\rangle &=& b_{3}\langle S_{0}\rangle+b_{2}\langle S_{0}^2\rangle  \\
\nonumber  \langle S_{0}S_{1}S_{2}^2\rangle &=& b_{3}\langle S_{0}S_{1}\rangle+b_{2}\langle S_{1}S_{0}^2\rangle \\
\nonumber  \langle S_{0}S_{1}^2S_{2}^2\rangle &=& b_{3}\langle S_{0}S_{1}^2\rangle+b_{2}\langle S_{1}^2S_{0}^2\rangle \\
\nonumber  \langle S_{1}S_{2}S_{3}^2\rangle &=& b_{1}\langle
S_{1}S_{2}\rangle
+b_{2}\langle S_{0}S_{1}S_{2}\rangle+(b_{3}-b_{1})\langle S_{1}S_{2}S_{0}^2\rangle  \\
\nonumber  \langle S_{1}S_{2}^2S_{3}^2\rangle &=& b_{1}\langle
S_{1}S_{2}^2\rangle
+b_{2}\langle S_{0}S_{1}S_{2}^2\rangle+(b_{3}-b_{1})\langle S_{1}S_{2}^2S_{0}^2\rangle \\
\nonumber  \langle S_{1}^2S_{2}^2S_{3}^2\rangle &=& b_{1}\langle
S_{1}^2S_{2}^2\rangle
+b_{2}\langle S_{0}S_{1}^2S_{2}^2\rangle+(b_{3}-b_{1})\langle S_{1}^2S_{2}^2S_{0}^2\rangle \\
\nonumber  \langle (S_{0}^{z})^2\rangle &=& r_{0}+3n_{1}\langle
S_{1}\rangle
+3(r_{1}-r_{0})\langle S_{1}^{2}\rangle+3r_{2}\langle S_{1}S_{2}\rangle  \\
\nonumber   &&+6(n_{2}-n_{1})\langle
S_{1}S_{2}^{2}\rangle+3(r_{0}-2r_{1}+r_{3})\langle
S_{1}^{2}S_{2}^{2}\rangle\\
\nonumber&&+n_{3}\langle\langle S_{1}S_{2}S_{3}\rangle+ 3(r_{4}-r_{2})\langle S_{1}S_{2}S_{3}^{2}\rangle  \\
\nonumber   &&+3(n_{1}-2n_{2}+n_{4})\langle S_{1}S_{2}^{2}S_{3}^{2}\rangle  \\
\nonumber&&+(-r_{0}+3r_{1}-3r_{3}+r_{5})\langle
S_{1}^{2}S_{2}^{2}S_{3}^{2}\rangle\\
\nonumber  \langle S_{1}S_{0}^2\rangle &=&(3r_{1}-2r_{0})\langle
S_{1}\rangle
+3n_{1}\langle S_{1}^2\rangle  \\
\nonumber && +(3r_{2}+3r_{0}-6r_{1}+3r_{3})\langle S_{1}S_{2}^2\rangle\\
&&\nonumber +6(n_{2}-n_{1})\langle
S_{1}^2S_{2}^2\rangle+n_{3}\langle
S_{1}S_{2}S_{3}^2\rangle\\
&&\nonumber+(-r_{0}+3r_{1}-3r_{2}-3r_{3}+3r_{4}+r_{5})\langle
S_{1}S_{2}^2S_{3}^2\rangle\\
&&\nonumber+3(n_{1}-2n_{2}+n_{4})\langle S_{1}^2S_{2}^2S_{3}^2\rangle\\
\nonumber  \langle S_{1}^2S_{0}^2\rangle &=& (3r_{1}-2r_{0})\langle
S_{1}^2\rangle
+3n_{1}\langle S_{1}\rangle \\
\nonumber&& +(3r_{2}+3r_{0}-6r_{1}+3r_{3})\langle
S_{1}^2S_{2}^2\rangle\\
\nonumber&&+6(n_{2}-n_{1})\langle S_{1}S_{2}^2\rangle\\
\nonumber&&+(3n_{1}-6n_{2}+n_{3}+3n_{4})\langle S_{1}S_{2}^2S_{3}^2\rangle\\
\nonumber&&+(-r_{0}+3r_{1}-3r_{2}-3r_{3}+3r_{4}+r_{5})\langle
S_{1}^2S_{2}^2S_{3}^2\rangle\\
\nonumber  \langle S_{1}S_{2}S_{0}^2\rangle &=& (r_{0}-3r_{1}
+3r_{2}+3r_{3})\langle S_{1}S_{2}\rangle\\
\nonumber&&+(-3n_{1}+6n_{2})\langle S_{1}S_{2}^2\rangle \\
\nonumber&& (-r_{0}+3r_{1}-3r_{2}-3r_{3}+3r_{4}+r_{5})\langle
S_{1}S_{2}S_{3}^2\rangle \\
\nonumber&&+(3n_{1}-6n_{2}+n_{3}+3n_{4})\langle
S_{1}S_{2}^2S_{3}^2\rangle
\end{eqnarray}
\begin{eqnarray}\label{eqA7}
\nonumber  \langle S_{1}S_{2}^2S_{0}^2\rangle &=&(r_{0}-3r_{1}
+3r_{2}+3r_{3})\langle S_{1}S_{2}^2\rangle\\
\nonumber&&+(-3n_{1}+6n_{2})\langle S_{1}S_{2}\rangle \\
\nonumber&& (-r_{0}+3r_{1}-3r_{2}-3r_{3}+3r_{4}+r_{5})\langle
S_{1}S_{2}^2S_{3}^2\rangle \\
\nonumber&&+(3n_{1}-6n_{2}+n_{3}+3n_{4})\langle S_{1}S_{2}S_{3}^2\rangle\\
\nonumber \langle S_{1}^{2}S_{2}^{2}S_{0}^{2}\rangle
\nonumber&=&({r_0}-3{r_1}+3{r_2}+3{r_3})\langle
S_{1}^{2}S_{2}^{2}\rangle\\
\nonumber&&+(-3{n_1}+6{n_2})\langle
{S_1}S_{2}^{2}\rangle\\
\nonumber&&+(3{n_1}-6{n_2}+{n_3}+3{n_4})\langle
{S_1}S_{2}^{2}S_{3}^{2}\rangle\\
\nonumber&&+(-{r_0}+3{r_1}-3{r_2}-3{r_3}+3{r_4}+{r_5})\langle
S_{1}^{2}S_{2}^{2}S_{3}^{2}\rangle\\
\nonumber \langle S_{0}^{x}\rangle&=&p_{0}+3c_{1}\langle S_{1}
\rangle+3(p_{1}-p_{0})\langle S_{1}^{2} \rangle+3p_{2}\langle
S_{1}S_{2}
\rangle \\
\nonumber& & +6(c_{2}-c_{1})\langle S_{1}S_{2}^{2}
\rangle+3(p_{0}-2p_{1}+p_{3})\langle S_{1}^{2}S_{2}^{2}
\rangle \\
\nonumber& &+c_{3}\langle S_{1}S_{2}S_{3}
\rangle+3(p_{4}-p_{2})\langle S_{1}S_{2}S_{3}^{2}
\rangle\\
\nonumber& &+3(c_{1}-2c_{2}+c_{4})\langle S_{1}S_{2}^{2}S_{3}^{2}
\rangle\\
\nonumber& &+(-p_{0}+3p_{1}-3p_{3}+p_{5})\langle
S_{1}^{2}S_{2}^{2}S_{3}^{2}
\rangle\\
\nonumber \langle (S_{0}^{x})^2\rangle&=&v_{0}+3\mu_{1}\langle S_{1}
\rangle+3(v_{1}-v_{0})\langle S_{1}^{2} \rangle+3v_{2}\langle
S_{1}S_{2}
\rangle \\
\nonumber& & +6(\mu_{2}-\mu_{1})\langle S_{1}S_{2}^{2}
\rangle+3(v_{0}-2v_{1}+v_{3})\langle S_{1}^{2}S_{2}^{2}
\rangle \\
\nonumber& &+\mu_{3}\langle S_{1}S_{2}S_{3}
\rangle+3(v_{4}-v_{2})\langle S_{1}S_{2}S_{3}^{2}
\rangle\\
\nonumber& &+3(\mu_{1}-2\mu_{2}+\mu_{4})\langle
S_{1}S_{2}^{2}S_{3}^{2}
\rangle\\
& &+(-v_{0}+3v_{1}-3v_{3}+v_{5})\langle S_{1}^{2}S_{2}^{2}S_{3}^{2}
\rangle
\end{eqnarray}

\newpage
\section*{References}


\begin{thebibliography}{99}

\bibitem{deGennes} P. G. de Gennes, Solid State Commun. 1 (1963) 132.

\bibitem{elliot} R. J. Elliot, G. A. Gehring, A. P. Malogemoff, S.
R. P. Smith, N. S. Staude, R. N. Tyte, J. Phys. C4 L (1971) 179.

\bibitem{wong} Y. L. Wong, B. Cooper, Phys. Rev. 172 (1968) 539.

\bibitem{fisher_rg} D. S. Fisher, Phys. Rev. Lett. 69 (1992) 534.

\bibitem{saber_eft} A. Saber, A. Ainane, F. Dujardin, M. Saber, B.
St\'{e}b\'{e}, J. Phys.: Condens. Matter 11 (1999) 2087.

\bibitem{sarmento_eft} E.F. Sarmento, I.P. Fittipaldi, T. Kaneyoshi,
J. Magn. Magn. Mater. 104-107 (1992) 233.

\bibitem{Saber} T. Bouziane, M. Saber J. Magn. Magn. Mater. 321
(2009) 17.

\bibitem{saxena_cvm} V. K. Saxena, Phys. Rev. B 27 (1983) 6884.

\bibitem{saxena_mf} V. K. Saxena, Phys. Lett. A 90 (1982) 71.

\bibitem{canko_pa} O. Canko, E. Albayrak, M. Keskin, J. Magn. Magn.
Mater. 294 (2005) 63.

\bibitem{creswick_mc} R. J. Creswick, H. A. Farach, J. M. Knight, C.
P. Poole Jr, Phys. Rev. B 38 (1988) 4712.

\bibitem{Jiang1} X. F. Jiang, J. L. Li, J. L. Zhong, C. Z. Yang,
Phys. Rev. B 47 (1993) 827.

\bibitem{Htoutou2} K. Htoutou, A. Oubelkacem, A. Ainane, M. Saber,
J. Magn. Magn. Mater. 288 (2005) 259.

\bibitem{Jiang2} X. F. Jiang, J. Magn. Magn. Mater. 134 (1994) 167.

\bibitem{Htoutou1} K. Htoutou, A. Benaboud, A. Ainane, M. Saber,
Physica A 338 (2004) 479.

\bibitem{Htoutou3} K. Htoutou, A. Ainane, M. Saber, J. J. de Miguel,
Physica A 358 (2005) 184.

\bibitem{wei_jiang1} W. Jiang, L. Q. Guo, G. Wei, A. Du, Physica B
307 (2001) 15.

\bibitem{wei_jiang2} W. Jiang, G. Wei, Z. H. Xin, Phys. Stat. Sol. B
225 (2001) 215.

\bibitem{Miao} H. Miao, G. Wei, J. Liu, J. Geng, J. Magn. Magn.
Mater. 321 (2009) 102.

\bibitem{polat} H. Polat, \"{U}. Ak{\i}nc{\i}, \.{I}. S\"{o}kmen, Phys. Status. Solidi B 240 (2003) 189.

\bibitem{canpolat} Y. Canpolat, A. Torg\"{u}rs\"{u}l, H. Polat, Phys. Scr. 76 (2007) 597.

\bibitem{yuksel} Y. Y\"{u}ksel, \"{U}. Ak{\i}nc{\i}, H. Polat, Phys. Scr. 79 (2009) 045009.

\bibitem{Barreto} F. C. S\'{a}Barreto, I. P. Fittipaldi, B.
Zeks, Ferroelectrics 39 (1981) 1103.

\bibitem{decoupling} I. Tamura, T. Kaneyoshi, Prog. Theor. Phys. 66 (1981)
1892.

\bibitem{Huang} K. Huang, Statistical Mechanics, Wiley Press, New York (1963).

\end{thebibliography}
\end{document}